# Incoherent Interference

Wang Kaige
*Department of Physics, Applied Optics Beijing Area Major Laboratory,*
*Beijing Normal University, Beijing 100875, China*
(Email: wangkg@bnu.edu.cn)

**Abstract**

From common knowledge, the occurrence of optical interference requires a coherent light source. The original meaning of coherence was attributed to the ability of light to exhibit interference phenomena. There are two types of coherence: temporal coherence and spatial coherence. For spatial interference to occur, usually both temporal and spatial coherence conditions have to be satisfied. Holography is one of the most interesting applications of spatial interference. In the first holography experiment in 1948, when the laser had not yet been invented, Dennis Gabor had to use a pinhole of 3 microns in diameter to improve the spatial coherence of the high-pressure mercury lamp. In the 1960s, however, incoherent holography techniques were developed to produce holograms of an object which is either self-luminous or is illuminated by spatially incoherent light. Two kinds of interference mechanism in holography are not the same. Soon after lasers were invented, coherent interference becomes the mainstream in the optical technology.

In the last decade of last century, theoretical and experimental studies have shown that a quantum light source—two-photon entangled state generated by spontaneous parametric down-conversion (SPDC) exhibits peculiar interference effects, such as sub-wavelength interference, nonlocal double-slit interference, ghost interference and diffraction, and ghost imaging, which were observed by the two-photon coincidence measurement. These effects were illustrated by the quantum interference theory of entangled photon pairs. But just at the beginning of the present century, it has been reported that all these effects can exist in the intensity correlation of thermal light. This is reminiscent of the Hanbury Brown and Twiss intensity interferometer for measuring the stellar diameter in 1956. The similarity in the second-order interference for quantum and classical sources caused controversy. There is one thing in common: both sources are spatially incoherent ones.

Since 2009, the experimental reports showed that spatial incoherent source can also perform the first-order interference which has different interference pattern from coherent pattern in the same setup. The main properties of the first-order incoherent interference are similar to that of the second-order one with thermal light. Both first-order and second-order incoherent interference show colorful phenomena, such as ghost interference and imaging, subwavelength interference, and nonlocal interference etc., that is never found in coherent interference.

Although coherent interference dominates the main part in optics, all sorts of the incoherent interference schemes can stimulate new physics and lead to various applications for when a coherent source is not available. In this article we review theoretical descriptions and experimental schemes of incoherent interference with both classical thermal light and quantum entangled light sources.

**CONTENTS**

## I.  Introduction

### 1.  Interference is a statistical concept in optics

It is well known that interference results from the nature of waves. When two or more waves are superposed, the intensity distribution can no longer in general be equal to the intensity superposition of each individual wave. The intensity distribution in the region of superposition is found to vary from point to point between maxima, which exceed the sum of the intensity in the waves, and minima, which may be zero. This phenomenon is called interference. Let us consider optical waves as an example. Assume two optical beams designated by electric fields $E_i(\mathbf{x})$ ($i$=1,2), where $\mathbf{x}$ is the position across the beam, then the intensity distribution of two superposed fields is given by

$$I(\mathbf{x}) = |E_1(\mathbf{x}) + E_2(\mathbf{x})|^2 = I_1(\mathbf{x}) + I_2(\mathbf{x}) + [E_1^*(\mathbf{x})E_2(\mathbf{x}) + \text{c.c.}], \qquad (1.1.1)$$

where $I_i(\mathbf{x}) = E_i^*(\mathbf{x})E_i(\mathbf{x})$ ($i$=1,2) is the intensity distribution of each individual beam. The interference seems to disappear only when the term $E_1^*(\mathbf{x})E_2(\mathbf{x})$ vanishes in a certain spatial area and/or a temporal duration. But actually both individual beams do not vanish there. And why cannot optical interference be observed everywhere?

Optics, along with mechanics, belongs to the oldest of the natural sciences, and experiments conducted by Mo Zi in China were recorded as early as in 400 BCE (Needham et al., 2003). However, the first optical interference effects, the colours exhibited by thin films, were only reported in the 17[th] century (Born and Wolf, 1999). This is because optical interference requires coherent light, which is hard to realize under natural conditions.

Light in nature is filled with fluctuations. Usually, the fluctuation is so fast it cannot be detected by the human eye. What we see is the average intensity of Eq. (1.1.1). Thus the interference term should be defined by

$$\left\langle E_1^*(\mathbf{x})E_2(\mathbf{x}) \right\rangle + \text{c.c.}, \qquad (1.1.2)$$

where $\langle \ \rangle$ denotes ensemble (or time) averaging. When the interference term of Eq. (1.1.2) vanishes, we say that interference does not occur, although the "instantaneous" interference term of Eq. (1.1.1) does not vanish. The definition of Eq. (1.1.2) incorporates the statistical property of light, i.e. the statistical superposition of Eq. (1.1.1) for optical waves. Moreover, for quantized optical fields, the interference definition of Eq. (1.1.2) involving the ensemble average of quantum state must be employed.

### 2.  Temporal and spatial coherence

It is generally regarded that coherence must be satisfied for interference to occur. Let us first recall our basic knowledge about coherence and interference in classical optics. The original meaning of coherence was attributed to the ability of light to exhibit interference phenomena. There are two types of coherence: temporal and spatial coherence. Temporal coherence measures the ability of two beams with a relative time delay to form interference fringes. Interference in a Michelson interferometer refers to temporal coherence. However, spatial coherence reflects the ability of a beam to interfere with another spatially shifted (but not delayed) beam (version across the beam). Young's double-slit experiment is an example which involves spatial coherence.

Coherence relies on the statistical properties of light. The optical field emitted from any light source in nature has its own random phase, which fluctuates in time. The average time during which the phase does not vary is called the coherence time. If the two fields participating in the interference, $E_1$ and $E_2$ in Eq. (1.1.2), have independent random phases, no interference will occur. When the spatial interference of Eq. (1.1.2) is considered, the temporal coherence must be taken into account. Therefore, the two fields that cause the interference must come from the same source, and the difference in their transit times from the source to the screen must be less than the coherence time of the source. Before the laser was invented, all natural light sources were thermal light, with a coherence time of less than $10^{-10}$ sec. As a result, it was not easy to observe interference in nature.

Spatial interference refers to the phenomenon where interference fringes appear in the plane perpendicular to the beam. Its occurrence requires not only temporal coherence but also, in general, spatial coherence. Spatial coherence is a measure of the phase correlation in the cross-section of the beam, and relates to the coherent area in the wavefront where the phase profile is well defined. For example, in Young's double-slit experiment, interference fringes can only be seen when the two slits are located within the coherent area of the wavefront which illuminates the slits.

In the early days when coherent sources were unavailable, interference experiments were carried out with a large area thermal source restricted by a pinhole aperture, which can improve the spatial coherence. Holography is one of the most important applications of spatial interference. In the first holography experiment, Dennis Gabor stated that (Gabor, 1972) "The best compromise between coherence and intensity was offered by the high-pressure mercury lamp, which had a coherence length of only 0.1 millimeter, ... But in order to achieve spatial coherence, we had to illuminate, with one mercury line, a pinhole 3 microns in diameter." The pinhole eventually reduced the power of the source and thus impeded the potential application of optical interferometric techniques such as holography. This barrier was overcome with the invention of the laser, whose intense and coherent beam was ideal for performing interference experiments.

Assuming the two fields in Eq. (1.1.2) come from the same monochromatic source $E_0(\mathbf{x})$, they can be expressed as

$$E_j(\mathbf{x},t) = \int h_j(\mathbf{x},\mathbf{x}_0) E_0(\mathbf{x}_0, t_{0j}) d\mathbf{x}_0, \quad (j=1,2), \tag{1.2.1}$$

where $h_j(\mathbf{x},\mathbf{x}_0)$ is the impulse response functions for path $j$. The interference term is given by

$$\langle E_1^*(\mathbf{x},t) E_2(\mathbf{x},t) \rangle = \int h_1^*(\mathbf{x},\mathbf{x}_0) h_2(\mathbf{x},\mathbf{x}_0') \langle E_0^*(\mathbf{x}_0, t_{01}) E_0(\mathbf{x}_0', t_{02}) \rangle d\mathbf{x}_0 d\mathbf{x}_0'. \tag{1.2.2}$$

Let $\tau$ be the coherence time of the source. When the time coherence requirement $|t_{02} - t_{01}| < \tau$ has not been satisfied, $\langle E_0^*(\mathbf{x}_0, t_{01}) E_0(\mathbf{x}_0', t_{02}) \rangle = 0$. The interference will not occur, i.e. $\langle E_1^*(\mathbf{x},t) E_2(\mathbf{x},t) \rangle = 0$. Spatial coherence requires that the beam has a well-defined wavefront, such as a field emitted from a point source or a laser. When the source field satisfies both temporal and spatial coherence, the first-order field correlation function can be written as

$$\langle E_0^*(\mathbf{x}_0, t_{01}) E_0(\mathbf{x}_0', t_{02}) \rangle = E_0^*(\mathbf{x}_0, t_{01}) E_0(\mathbf{x}_0', t_{02}). \tag{1.2.3}$$

Accordingly, the interference term has the similar form

$$\langle E_1^*(\mathbf{x},t) E_2(\mathbf{x},t) \rangle = E_1^*(\mathbf{x},t) E_2(\mathbf{x},t). \tag{1.2.4}$$

This is called *coherent interference*. Most of the interference phenomena discussed in the literature satisfy the above equation. If the light is sufficiently intense so that quantum fluctuations can be disregarded, an interference pattern can be observed at any single moment.

3. **Is spatial coherence necessary for interference?**

The above analysis shows that temporal coherence is a necessary condition for interference. We will then naturally raise the question: is spatial coherence also necessary for interference? Let us consider the case when the source field satisfies temporal but not spatial coherence. For complete spatial incoherence, the beam fluctuates randomly and any two positions on its wavefront are statistically independent. The first-order field correlation function for the source field satisfying complete spatial incoherence is written as

$$\langle E_0^*(\mathbf{x}_0) E_0(\mathbf{x}_0') \rangle = I_0 \delta(\mathbf{x}_0 - \mathbf{x}_0'), \tag{1.3.1}$$

where, for simplicity, the intensity distribution at the source is assumed to be homogeneous value $I_0$. Substituting Eq. (1.3.1) into Eq. (1.2.2), we obtain

$$\left\langle E_1^*(\mathbf{x})E_2(\mathbf{x})\right\rangle = I_0 \int h_1^*(\mathbf{x},\mathbf{x}_0)h_2(\mathbf{x},\mathbf{x}_0)d\mathbf{x}_0. \tag{1.3.2}$$

At this stage, it is not clear if Eq. (1.3.2) is able to present an interference pattern.

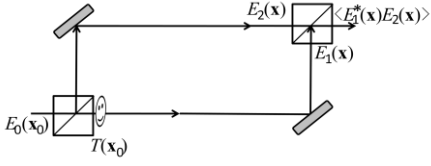

Fig. 1-1 Sketch of holographic interference.

Let us reconsider Gabor's original holographic scheme as shown in Fig.1-1 but the source satisfies spatial incoherence (1.3.1). In the paraxial approximation, the impulse response function for free propagation through a distance $z$ in vacuum is given by $H(\mathbf{x},\mathbf{x}_0;z,z)$, where the function $H$ is defined as

$$H(\mathbf{x},\mathbf{x}_0;z,Z) = \frac{k_0}{\mathrm{i}2\pi Z}\exp(\mathrm{i}k_0 z)\exp\left[\frac{\mathrm{i}k_0}{2Z}|\mathbf{x}-\mathbf{x}_0|^2\right], \tag{1.3.3}$$

and $k_0$ is the wave number in vacuum. As the wave propagates through a distance $l$ in a medium of refractive index $n$, we have $z=nl$ and $Z=l/n$. A transmitting object with transmittance $T(\mathbf{x})$ is placed in one path, close to the source. In order to satisfy temporal coherence, the two paths should have equal propagation lengths $z_1=z_2=z$, so $h_1(\mathbf{x},\mathbf{x}_0) = T(\mathbf{x}_0)H(\mathbf{x},\mathbf{x}_0;z,z)$ and $h_2(\mathbf{x},\mathbf{x}_0) = H(\mathbf{x},\mathbf{x}_0;z,z)$. Therefore, the interference term (1.3.2) yields

$$\left\langle E_1^*(\mathbf{x})E_2(\mathbf{x})\right\rangle \propto I_0 \int T(\mathbf{x}_0)d\mathbf{x}_0, \tag{1.3.4}$$

where the object information $T(\mathbf{x}_0)$ has been washed out completely although the interference term is not null in general. We can also prove that, no matter where the object is placed within the object path, Eq. (1.3.4) still holds provided the two paths of the interferometer have the same length. Perhaps this result brought about the misconception that spatial interference needs not only temporal coherence but also spatial coherence.

Since the first double-slit interference experiment, coherent interference has been the main tune in the history of optics, while only a few studies have considered incoherent interference. In this review article, we shall recall those interference effects where spatial coherence is not satisfied.

## II. Incoherent holography

In the 1960s, incoherent holography techniques were developed to produce holograms of an object which is either self-luminous or is illuminated by spatially incoherent light (Rogers, 1977; Goodman, 1996; Mertz and Young, 1962; Mertz, 1964; Lohmann, 1965; Stroke and Restrick, 1965; Winthrop and Worthington, 1965; Cochran, 1965; Cochran, 1966; Peters, 1966; Worthington, 1966; Montgomery, 1966; Bryndahl and Lohmann, 1968). In one method, the separate coherent reference beam in spatially coherent holography is replaced by a reference beam which is generated by the object itself. Each object point of the source generates its own interference pattern by some optical means, which uniquely encodes the position and intensity of the source. The holographic recording is built up by an incoherent superposition of the interference patterns originating from all points of the object. The use of the incoherent hologram technique is advantageous only in situations where spatially coherent light is not available.

Mertz and Young first proposed a way of generating a hologram (Mertz and Young, 1962), which could be used as a camera for x-ray astronomy. Each star as an object point generates its own Fresnel zone plate (FZP) pattern on the hologram. Later, each FZP mask acts as a lens in the reconstruction process and produces an image of the star.

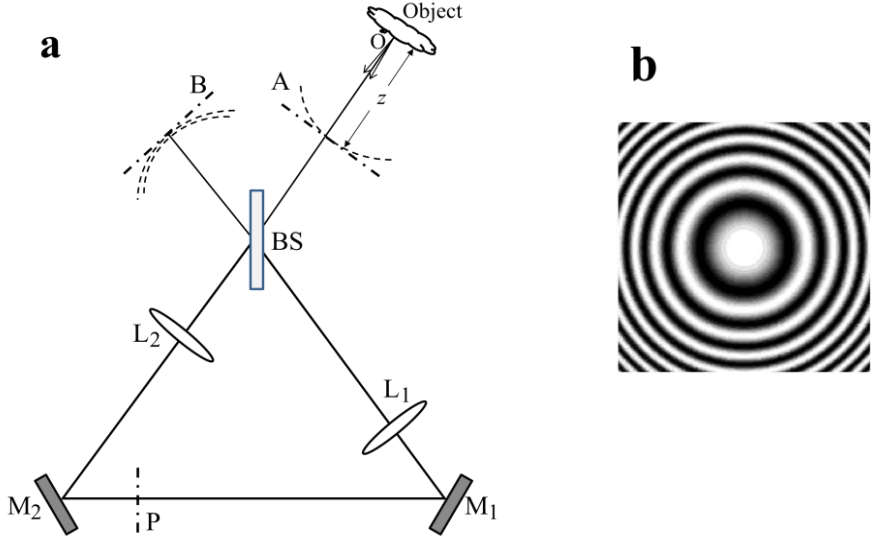

Fig. 2-1 (a) Sketch of a triangular interferometer. BS is a beamsplitter, $M_1$ and $M_2$ are two mirrors, $L_1$ and $L_2$ are two lenses with their focal planes located at A and B, respectively. (b) Fresnel zone plate (FZP) pattern.

To create an FZP, a spherical wave coming from each object point is split into two portions with different curvatures in some special optical schemes. These two portions are coherent since they come from the same source point. When they are recombined, the interference pattern has a FZP shape (Mertz, 1964; Lohmann, 1965; Stroke and Restrick, 1965; Cochran, 1965; Cochran, 1966; Peters, 1966). A typical scheme to form an FZP is shown in Fig. 2-1a, where the triangular interferometer (Mertz, 1964; Cochran, 1966) consists of a beamsplitter BS and two mirrors $M_1$ and $M_2$. The two lenses $L_1$ and $L_2$ with different focal lengths $f_1$ and $f_2$ are placed in the two arms of the interferometer. Planes A and B are the focal planes of lenses $L_1$ and $L_2$, respectively. The other focal planes of lenses $L_1$ and $L_2$ coincide at plane P. A beam, emitted from plane A, is reflected by BS, and travels clockwise around the interferometer and finally arrives at plane B. This forms an afocal optical system, in which the wavefront $E(x,y)$ in plane A will be imaged as $E(x/m_1,y/m_1)$ in plane B, where $m_1=-f_2/f_1$ is the magnification. At the same time, the beam is also transmitted through BS, and travels counterclockwise around the interferometer and finally arrives at plane B. Again the wavefront $E(x,y)$ in plane A will be imaged as $E(x/m_2,y/m_2)$ in plane B with the magnification $m_2=-f_1/f_2$. Therefore the triangle interferometer forms a double afocal system (Cochran, 1966), where a wavefront in plane A will generate two image-wavefronts, coherently superposed in plane B.

Consider now a point source of an incoherent object located at position O, which is at a distance $z$ from plane A. The corresponding wavefront in plane A can be expressed as $E_A(x,y)=E_0\exp[ik(x^2+y^2)/(2z)]$. After a double afocal system in the triangle interferometer, the superposed fields in plane B will be

$$E_B(x, y) = rE_A(x/m_1, y/m_1) + tE_A(x/m_2, y/m_2)$$
$$= rE_0 \exp\left[\frac{ik}{2z}\frac{x^2+y^2}{m_1^2}\right] + tE_0 \exp\left[\frac{ik}{2z}\frac{x^2+y^2}{m_2^2}\right], \tag{2.1}$$

where $r$ and $t$ are reflectivity and transmissivity of the amplitude for the beamsplitter. For a lossless beamsplitter satisfying $r^2 + t^2 = 1$, the intensity distribution is given by

$$I_B(x, y) = |E_0|^2 \left\{1 + 2rt \cos\left[\frac{k}{2F}(x^2+y^2)\right]\right\}, \tag{2.2}$$

where $F = \dfrac{zf_1^2 f_2^2}{f_1^4 - f_2^4}$. Equation (2.2) defines a FZP encoding point O of the incoherent object (see Fig. 2-1b). However, an adjacent point source produces an adjacent FZP. The hologram then consists of the incoherent addition of these zone plates.

One may ask a question: will the incoherent superposition of all the FZPs wash out the holographic information? To answer this question, we consider the general formula of incoherent holography. Let $E_0(\mathbf{x})$ be the field amplitude of incoherent object, satisfying

$$\langle E_0^*(\mathbf{x}_0) E_0(\mathbf{x}_0') \rangle = I_0(\mathbf{x}_0)\delta(\mathbf{x}_0 - \mathbf{x}_0'). \tag{2.3}$$

In incoherent holography, the light emitted or reflected from the object experiences two different paths, described by the impulse response functions $h_1(\mathbf{x},\mathbf{x}_0)$ and $h_2(\mathbf{x},\mathbf{x}_0)$. When the two portions of the fields rejoin, the interference term is written as

$$\langle E_1^*(\mathbf{x}) E_2(\mathbf{x}) \rangle = \int I_0(\mathbf{x}_0) h_1^*(\mathbf{x},\mathbf{x}_0) h_2(\mathbf{x},\mathbf{x}_0) d\mathbf{x}_0. \tag{2.4}$$

For the interferometer of Fig. 2-1a, the two impulse response functions for the clockwise and counterclockwise directions from plane A to plane B are given by

$$h_{ABj}(\mathbf{x},\mathbf{x}_1) = (1/m_j)\exp[i2k(f_1+f_2)]\delta^2(\mathbf{x}_1 - \mathbf{x}/m_j), \quad (j=1,2), \tag{2.5}$$

where $m_j$ is the magnification defined above. The two impulse response functions from the object to plane B can be obtained by

$$h_j(\mathbf{x},\mathbf{x}_0) = \int H(\mathbf{x}_1,\mathbf{x}_0;z,z)h_{ABj}(\mathbf{x},\mathbf{x}_1)d\mathbf{x}_1$$
$$= \frac{k}{i2\pi z m_j}\exp[ik(2f_1+2f_2+z)]\exp\left[\frac{ik}{2z}\left|\frac{\mathbf{x}}{m_j}-\mathbf{x}_0\right|^2\right], \quad (j=1,2). \tag{2.6}$$

According to Eq. (2.4), we arrive at a well-defined interference pattern

$$\langle E_1^*(\mathbf{x})E_2(\mathbf{x}) \rangle = \left(\frac{k}{2\pi z}\right)^2 \exp\left[\frac{ik}{2F}|\mathbf{x}|^2\right] \int I_0(\mathbf{x}_0)\exp\left[\frac{ik}{z}\left(\frac{f_2}{f_1}-\frac{f_1}{f_2}\right)\mathbf{x}\cdot\mathbf{x}_0\right]d\mathbf{x}_0, \tag{2.7}$$

which encodes the object $I_0(\mathbf{x}_0)$.

The other schemes of incoherent holography are based on the idea to create (really or virtually) two images from one object, which might be rotated and shifted laterally (Stroke and Restrick, 1965; Winthrop and Worthington, 1965; Worthington, 1966; Montgomery, 1966; Bryndahl and Lohmann, 1968). Corresponding pairs of image points are mutually coherent, and hence they can form interference fringes on the hologram. A typical scheme can be explained in Fig. 2-2.

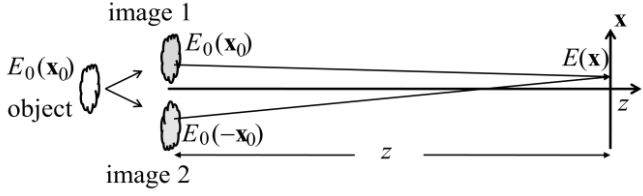

Fig. 2-2 Incoherent interference of two images from one object.

In Fig. 2-2, an object is described by the incoherent field $E_0(\mathbf{x}_0)$ satisfying Eq. (2.3). Then the object is copied to form two images, $E_0(\mathbf{x}_0)$ and $E_0(-\mathbf{x}_0)$. It is not necessary for the two images to be separated laterally. After travelling a free distance $z$, the superposed field $E(\mathbf{x})$ reads

$$E(\mathbf{x}) = \int H(\mathbf{x},\mathbf{x}_0;z,z)E_0(\mathbf{x}_0)d\mathbf{x}_0 + \int H(\mathbf{x},\mathbf{x}_0;z,z)E_0(-\mathbf{x}_0)d\mathbf{x}_0 . \qquad (2.8)$$

The corresponding intensity distribution can be derived as

$$\langle E^*(\mathbf{x})E(\mathbf{x})\rangle = \frac{k^2}{2\pi^2 z^2}\int I_0(\mathbf{x}_0)[1+\cos(2k\mathbf{x}\cdot\mathbf{x}_0/z)]d\mathbf{x}_0 . \qquad (2.9)$$

The above method was proposed in 1965 by Stroke and Restrick, who first realized lensless Fourier-transform holography with incoherent light (Stroke and Restrick, 1965).

The schemes proposed for incoherent holography can only record the intensity distribution of a fluorescent object. Moreover, each elementary fringe pattern is formed by two extremely tiny portions of the light, and the summation of many weak interference patterns results in a very large bias level in the hologram, much larger than that in coherent holography. Hence incoherent holography is appropriate only for objects with a low number of resolution elements (Goodman, 1996).

### III. Second-order interference (two-photon interference)

The interference of spatially incoherent light can also be carried out with intensity correlation measurement. In 1956 R. Hanbury Brown and R. Q. Twiss (HB-T) proposed an intensity interferometer to measure the angular diameter of a star (Hanbury Brown and Twiss, 1956). The interference appeared in their experiment through the intensity correlation, defined as the second-order interference, differently from the first-order interference based on the intensity itself. First-order and second-order interferences are also called one-photon and two-photon interferences, respectively.

1. **Hanbury Brown and Twiss intensity interferometer**

In astronomy, optical interferometry is used to measure the angular diameter of a star. The star can be regarded as a thermal light source satisfying the complete spatial incoherence of Eq. (1.3.1). When the star light arrives at Earth, the first-order field correlation function is written as

$$\langle E^*(\mathbf{x}_1)E(\mathbf{x}_2)\rangle = I_0\int C(\mathbf{x}_0)H^*(\mathbf{x}_1,\mathbf{x}_0;z,z)H(\mathbf{x}_2,\mathbf{x}_0;z,z)d\mathbf{x}_0 , \qquad (3.1.1)$$

where $C(\mathbf{x}_0) = \begin{cases} 1 & |\mathbf{x}_0| \leq D/2 \\ 0 & |\mathbf{x}_0| > D/2 \end{cases}$ is the circle function, satisfying $|C(\mathbf{x}_0)|^2 = C(\mathbf{x}_0)$, and $D$ is the star's diameter. In the far-field limit, we obtain

$$\left|\langle E^*(\mathbf{x}_1)E(\mathbf{x}_2)\rangle\right| \propto \left|\frac{2J_1(\pi Dd/(\lambda z))}{\pi Dd/(\lambda z)}\right|, \qquad (3.1.2)$$

where $J_1$ is the first-order Bessel function of the first kind, $\lambda$ is the wavelength, $z$ is the distance from the star to Earth, and $d = |\mathbf{x}_1 - \mathbf{x}_2|$ is the distance between two positions on the earth. When $d = d_c$ the coherence length is defined for $J_1 = 0$, we can obtain the angular diameter of the star $\theta \equiv D/z = 1.22\lambda/d_c$.

The coherence length $d_c$ can be measured by the Michelson stellar interferometer. As shown in Fig. 3-1a, two small mirrors M1 and M2, separated by a distance $d$, reflect the star light into the Michelson interferometer. By observation of the visibility of the interference fringes in the interferometer, we can find the coherence length $d_c$.

The intensity interferometer is shown in Fig. 3-1b (Hanbury Brown and Twiss, 1956). Light from a star is received on two separate photo electric detectors, and produces output currents. These currents fluctuate and an intensity interferometer relies on the fact that the fluctuations are partially correlated. The two fluctuating currents are multiplied together in the linear multiplier, and this product, or correlation, is proportional to the square of the degree of coherence of the light at the two detectors

$$\langle \Delta I(\mathbf{x}_1)\Delta I(\mathbf{x}_2)\rangle = \langle I(\mathbf{x}_1)I(\mathbf{x}_2)\rangle - \langle I(\mathbf{x}_1)\rangle\langle I(\mathbf{x}_2)\rangle = \left|\langle E^*(\mathbf{x}_1)E(\mathbf{x}_2)\rangle\right|^2. \qquad (3.1.3)$$

We can see that the intensity interferometer has a simpler optical configuration than the Michelson interferometer. It is clear that its interference effect cannot be regarded as coherence interference like that in the Michelson interferometer, where the interference fringes appear instantly. As a matter of fact, each detector in the intensity interferometer monitors irregular intensity fluctuation, while the coherence information appears only in the statistical summation of intensity fluctuation products of the two detectors.

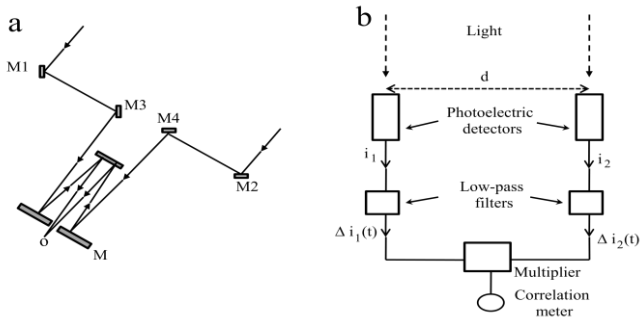

Fig. 3-1 (a) Michelson stellar interferometer; (b) Stellar intensity interferometer. [Adapted from (Hanbury Brown, 1974).]

We have already defined the first-order interference and the second-order interference at the beginning of the present section. Before the HB-T experiment, optical interference effects were observed through the intensity distribution and all pertained to the first-order interference. The HB-T experiment is regarded as the birth of quantum optics. In 1963 Roy Glauber established the quantum coherence theory of light (Glauber, 1963). According to this theory, the second-order spatial coherence is defined by both

$$\langle E^*(\mathbf{x})E(\mathbf{x}')\rangle = E^*(\mathbf{x})E(\mathbf{x}'), \tag{3.1.4a}$$

and

$$\langle E^*(\mathbf{x}_1)E^*(\mathbf{x}_2)E(\mathbf{x}_2')E(\mathbf{x}_1')\rangle = E^*(\mathbf{x}_1)E^*(\mathbf{x}_2)E(\mathbf{x}_2')E(\mathbf{x}_1'). \tag{3.1.4b}$$

When an optical field satisfies the second-order spatial coherence condition, then

$$\langle I(\mathbf{x}_1)I(\mathbf{x}_2)\rangle = I(\mathbf{x}_1)I(\mathbf{x}_2), \tag{3.1.5}$$

where there is neither intensity correlation nor intensity fluctuation (set $\mathbf{x}_1=\mathbf{x}_2$). Moreover, Eqs. (3.1.4) and (3.1.5) manifest the fact that when the field satisfies second-order coherence, the second-order interference may not contain more information than what the first-order coherence has provided. *We notice that for the second-order interference effects that have been studied so far in the literature, first-order interference in the same configuration does not exist; that is, first-order spatial coherence is not satisfied. In this sense these second-order interference effects (HB-T intensity interferometer and others as shown below) can be regarded as incoherent interference.*

2. **Fano model: interference from two phase-independent sources**

In the photon picture, the first-order and second-order interferences are referred to as the one-photon and two-photon effects, respectively. As a matter of fact, when only one photon passes through the Michelson interferometer each time, the interference pattern still appears after a long accumulation time. So the first-order interference is usually called one-photon interference and follows Dirac's famous statement "Each photon then interferes only with itself" (Dirac, 1930). Obviously, this is not the case for the intensity correlation measurement in the HB-T intensity interferometer, where at least two photons jointly interfere, and the interference relies on two-photon correlation.

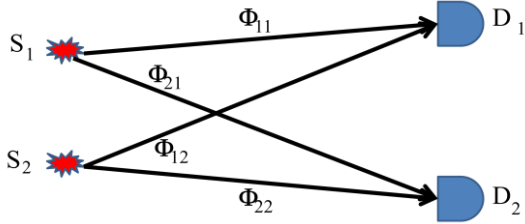

Fig. 3-2 Fano model. $S_1$ and $S_2$ are two sources, and $D_1$ and $D_2$ are two detectors. $\Phi_{ji}$ is the impulse response function from $S_i$ to $D_j$.

A few years after the HB-T experiment, U. Fano proposed a very simple quantum model to illustrate the HB-T effect (Fano, 1961). As shown in Fig. 3-2, $S_1$ and $S_2$ are two independently excited atoms, which emit a single photon one at a time, and $D_1$ and $D_2$ are two detector atoms. As stated by Fano, this system cannot produce any first-order interference phenomenon, which implies incoherence. Fano nicely considered two-particle probability amplitudes which are not affected by the random phases of the source atoms.

Actually, Fano model provides a more general understanding of the second-order interference in both classical and quantum systems if we consider that the two point sources are phase-independent. In a general description, the field at detector $D_j$ is the superposition of the fields emitted from the two sources:

$$E(D_j) = E_1(D_j) + E_2(D_j) \qquad (j=1,2), \tag{3.2.1}$$

where $E_i(D_j) = E_{0i}\Phi_{ji}$ is the field from source $S_i$ to detector $D_j$, $E_{0i}$, the field of source $S_i$ and $\Phi_{ji}$, the impulse response function from $S_i$ to $D_j$. The corresponding intensity is written as

$$I(D_j) = |E_1(D_j)|^2 + |E_2(D_j)|^2 + [E_1^*(D_j)E_2(D_j) + \text{c.c.}] \quad (j=1,2). \tag{3.2.2}$$

Since the two sources have independent phases, the average intensity at each detector does not display first-order interference, i.e. $\langle E_1^*(D_j)E_2(D_j)\rangle=0$.

The intensity correlation measurement of the two detectors in Fano model is obtained to be

$$\langle I(D_1)I(D_2)\rangle = \langle |E_1(D_1)|^2 |E_1(D_2)|^2\rangle + \langle |E_2(D_1)|^2 |E_2(D_2)|^2\rangle$$
$$+ \langle |E_1(D_1)|^2 |E_2(D_2)|^2\rangle + \langle |E_2(D_1)|^2 |E_1(D_2)|^2\rangle + [\langle E_1^*(D_1)E_2(D_1)E_1(D_2)E_2^*(D_2)\rangle + \text{c.c.}]$$
$$= |E_{01}|^4 |\Phi_{11}\Phi_{21}|^2 + |E_{02}|^4 |\Phi_{12}\Phi_{22}|^2$$
$$+ |E_{01}|^2|E_{02}|^2 [|\Phi_{11}\Phi_{22}|^2 + |\Phi_{12}\Phi_{21}|^2 + (\Phi_{11}^*\Phi_{22}^*\Phi_{12}\Phi_{21} + \text{c.c.})]. \tag{3.2.3}$$

We can see that there is a second-order interference term consisting of $E_1^*(D_1)E_2(D_1)$ and $E_1(D_2)E_2^*(D_2)$, in which the random phases of the two sources have been exactly eliminated in the product. However, due to the independence of two sources we can write

$$\langle E_1^*(D_1)E_2(D_1)E_1(D_2)E_2^*(D_2)\rangle = \langle E_1^*(D_1)E_1(D_2)\rangle\langle E_2(D_1)E_2^*(D_2)\rangle. \tag{3.2.4}$$

The second-order interference term is factorized into two first-order field correlation functions of two positions. The latter do not vanish provided the temporal coherence of each source is satisfied. The above equation (3.2.4) also tells us that when the two detectors are in the same place, the second-order interference term cannot give rise to spatial interference.

Moreover, the first-order correlation function of the superposed field is the sum of those of the source fields:

$$\langle E^*(D_1)E(D_2)\rangle = \langle E_1^*(D_1)E_1(D_2)\rangle + \langle E_2^*(D_1)E_2(D_2)\rangle, \tag{3.2.5}$$

and can be related to the second-order interference in the form of

$$|\langle E^*(D_1)E(D_2)\rangle|^2 = |\langle E_1^*(D_1)E_1(D_2)\rangle|^2 + |\langle E_2^*(D_1)E_2(D_2)\rangle|^2$$
$$+ \left[\langle E_1^*(D_1)E_2(D_1)E_1(D_2)E_2^*(D_2)\rangle + \text{c.c.}\right]. \tag{3.2.6}$$

Hence we obtain the correlation of the intensity fluctuations at the two detectors to be

$$\langle I(D_1)I(D_2)\rangle - \langle I(D_1)\rangle\langle I(D_2)\rangle = |\langle E^*(D_1)E(D_2)\rangle|^2$$
$$+ \langle |E_1(D_1)|^2 |E_1(D_2)|^2\rangle - \langle |E_1(D_1)|^2\rangle\langle |E_1(D_2)|^2\rangle - |\langle E_1^*(D_1)E_1(D_2)\rangle|^2$$
$$+ \langle |E_2(D_1)|^2 |E_2(D_2)|^2\rangle - \langle |E_2(D_1)|^2\rangle\langle |E_2(D_2)|^2\rangle - |\langle E_2^*(D_1)E_2(D_2)\rangle|^2. \tag{3.2.7}$$

When the two sources are thermal light for which the intensity correlation relation of Eq. (3.1.3) is valid, the superposition field also obeys Eq. (3.1.3).

The above general description of Fano model concludes the following points:

(i) For a superposed field of phase-independent sources, second-order interference may occur while even when the first-order interference disappears, no matter whether the sources are laser beams or thermal light, classical or quantum.

(ii) In Fano model the second-order interference pattern can be observed through the intensity correlation measurement at separate spatial positions.

(iii) The second-order interference can be factorized into two first-order field correlations. Thus the necessary condition for second-order interference is temporal coherence for each source. This implies an important fact: *although the second-order interference needs at least two photons, each photon still interferes with itself*.

(iv) The intensity correlation relation (3.1.3) is valid for each independent thermal light source, and it is also valid for any sum of them.

In the following subsection, we will discuss the quantum interference when Fano model is applied to a two-photon entangled source.

3. **Theoretical description of spatial coherence and correlation for single-photon and two-photon states**

About thirty years after the HB-T intensity interferometer appeared, a number of quantum interference effects with a two-photon entangled state were studied (Hong et al., 1987; Rarity et al., 1990; Belinskii and Klyshko, 1994; Pittman et al., 1995; Strekalov et al., 1995; Fonseca et al., 1999; Boto et al., 2000; D'Angelo et al., 2001; Song et al., 2013) such as Hong-Ou-Mandel (HOM) interference (Hong et al., 1987; Mandel, 1999; Wang, 2006), two-photon interference in a Mach-Zehnder interferometer (Rarity et al., 1990), ghost diffraction, ghost imaging, quantum holography (Belinskii and Klyshko, 1994; Pittman et al., 1995; Strekalov et al., 1995; Song et al., 2013), and subwavelength interference (Fonseca et al., 1999a; Boto et al., 2000; D'Angelo et al., 2001; Cao and Wang, 2004; Cao et al., 2005a), and so forth. Similar to the intensity correlation in the HB-T effect, these two-photon interference effects are observed through coincidence measurement of two photons, rather than one photon detection. But these effects relied on a quantum entangled source.

How is the coherence of quantum states manifested for one photon and two photons? Assume a single-photon pure state of a monochromatic field in the superposition of spatial modes

$$|\psi_s\rangle = \int C(p) a^\dagger(p) |0\rangle \mathrm{d}p = \int C(x) a^\dagger(x) |0\rangle \mathrm{d}x, \qquad (3.3.1)$$

where $p$ and $x$ are momentum and position, respectively. Function $C(x)$ and the creation operator $a^\dagger(x)$ in the position space are Fourier transforms of $C(p)$ and $a^\dagger(p)$ in the momentum space, respectively. In general, the single-photon pure state (3.3.1) is a coherently superposed spatially multimode state. The optical field operator for a monochromatic beam of frequency $\omega$ is given by

$$E^{(+)}(x) = \frac{\mathrm{i}}{(2\pi)^{3/2}} \left(\frac{\hbar\omega}{2\varepsilon_0}\right)^{1/2} a(x) = \frac{\mathrm{i}}{(2\pi)^{3/2}} \left(\frac{\hbar\omega}{2\varepsilon_0}\right)^{1/2} \int a(p) \exp(\mathrm{i}xp) \mathrm{d}p. \qquad (3.3.2)$$

The one-photon amplitude (or one-photon wavepacket) is defined as $\langle 0|E^{(+)}(x)|\psi_s\rangle$ (Scully and Zubairy, 1997). Using Eqs. (3.3.1) and (3.3.2), we obtain

$$\langle 0|E^{(+)}(x)|\psi_s\rangle \propto C(x), \qquad (3.3.3)$$

which is a well-defined optical field in the C-number form for the single photon. As a result, the first-order correlation function for the single-photon pure state can be factorized as

$$\langle \psi_s|E^{(-)}(x_1)E^{(+)}(x_2)|\psi_s\rangle = \langle \psi_s|E^{(-)}(x_1)|0\rangle\langle 0|E^{(+)}(x_2)|\psi_s\rangle \propto C^*(x_1)C(x_2). \qquad (3.3.4)$$

This equation is similar to Eq. (1.2.4) in its mathematical form. In this sense, any single-photon pure state (3.3.1) of a monochromatic field satisfies first-order spatial coherence. Although $\langle 0|E^{(+)}(x)|\psi_s\rangle$ is not observable, $\langle \psi_s|E^{(-)}(x)E^{(+)}(x)|\psi_s\rangle$ is proportional to the intensity distribution in the ensemble average that the single-photon detector records. However, it should be pointed out that the coherent interference Eq. (1.2.4) for classical fields has an explicit physical meaning, in that any single measurement is meaningful. However, the quantum expression of Eq. (3.3.4) does not have such a meaning because of quantum fluctuation.

A single-photon multimode mixed state of monochromatic field is expressed by the density operator

$$\rho_s = \sum_p Q(p)|1(p)\rangle\langle 1(p)|, \qquad (3.3.5)$$

where $Q(p)$ is the probability density of momentum $p$. The mixed single-photon state is an incoherent superposition of spatial modes. We can readily obtain

$$\langle a^\dagger(p_1)a(p_2)\rangle = \text{Tr}[\rho_s a^\dagger(p_1)a(p_2)] = Q(p_1)\delta(p_1-p_2). \qquad (3.3.6)$$

The first-order field correlation function for the single-photon mixed state is now

$$\langle E^{(-)}(x_1)E^{(+)}(x_2)\rangle \propto \tilde{Q}(x_1-x_2), \qquad (3.3.7)$$

where $\tilde{Q}(x)$ is the Fourier transform of $Q(p)$. When $Q(p)$ is a flat spectrum, the single-photon multimode mixed state is totally spatially incoherent:

$$\langle E^{(-)}(x_1)E^{(+)}(x_2)\rangle \propto \delta(x_1-x_2). \qquad (3.3.8)$$

However, for both pure and mixed single-photon states, the value of any higher-order correlation function is zero.

A two-photon pure state can be expressed as

$$|\psi_t\rangle = \int dp_1 dp_2 C(p_1,p_2)a_1^\dagger(p_1)a_2^\dagger(p_2)|0\rangle = \int dx_1 dx_2 C(x_1,x_2)a_1^\dagger(x_1)a_2^\dagger(x_2)|0\rangle. \qquad (3.3.9a)$$

In general, two photons in the state can be in different modes, such as different polarizations or frequencies. If frequency entanglement is present, the two-photon state has the form of

$$|\psi_t\rangle = \int d\omega_1 d\omega_2 C(\omega_1,\omega_2)a_1^\dagger(\omega_1)a_2^\dagger(\omega_2)|0\rangle, \qquad (3.3.9b)$$

where $a_1^\dagger$ and $a_2^\dagger$ define, for example, two spatial modes or two polarizations.

Similarly, the two-photon amplitude (or two-photon wavepacket) can be defined to be $\langle 0|E_1^{(+)}(x_1)E_2^{(+)}(x_2)|\psi_t\rangle$, where $E_i^{(+)}$ is the field operators for mode $i$. With Eqs. (3.3.2) and (3.3.9a) we arrive at

$$\langle 0|E_1^{(+)}(x_1)E_2^{(+)}(x_2)|\psi_t\rangle \propto C(x_1,x_2). \qquad (3.3.10)$$

While one-photon amplitude can correspond to classical field amplitude, the two-photon amplitude for the entanglement case has no classical correspondence.

When the two-photon amplitude can be factorized to

$$C(x_1,x_2) = C_1(x_1)C_2(x_2), \qquad (3.3.11)$$

the two-photon state (3.3.9) consists of two independent single-photon states

$$|\psi_t\rangle = |\psi_{s1}\rangle|\psi_{s2}\rangle = \int C_1(x_1)a_1^\dagger(x_1)|0\rangle dx_1 \cdot \int C_2(x_2)a_2^\dagger(x_2)|0\rangle dx_2. \qquad (3.3.12)$$

Accordingly, the two-photon amplitude is decomposed into two independent single-photon amplitudes

$$\langle 0|E_1^{(+)}(x_1)E_2^{(+)}(x_2)|\psi_t\rangle = \langle 0|E_1^{(+)}(x_1)|\psi_{s1}\rangle \times \langle 0|E_2^{(+)}(x_2)|\psi_{s2}\rangle \propto C_1(x_1)C_2(x_2). \qquad (3.3.13)$$

However, when the factorization (3.3.13) is not possible, the two-photon state becomes a quantum entangled state. In particular, for perfect entanglement, we have

$$\langle 0|E_1^{(+)}(x_1)E_2^{(+)}(x_2)|\psi_t\rangle \propto C(x_1,x_2) = \delta(x_1 - x_2).  \quad (3.3.14)$$

We consider the first-order and second-order correlation functions for the two-photon state (3.3.9a). The first-order correlation functions for the two fields are calculated to be

$$\langle \psi_t|E_1^{(-)}(x_1)E_1^{(+)}(x_2)|\psi_t\rangle \propto \int C^*(x_1,x)C(x_2,x)\mathrm{d}x, \quad (3.3.15a)$$

$$\langle \psi_t|E_2^{(-)}(x_1)E_2^{(+)}(x_2)|\psi_t\rangle \propto \int C^*(x,x_1)C(x,x_2)\mathrm{d}x, \quad (3.3.15b)$$

$$\langle \psi_t|E_1^{(-)}(x_1)E_2^{(+)}(x_2)|\psi_t\rangle = \langle \psi_t|E_2^{(-)}(x_1)E_1^{(+)}(x_2)|\psi_t\rangle = 0. \quad (3.3.16)$$

The first-order field correlation exists only within the same mode. However, Eq. (3.3.16) shows that two photons in different modes cannot display first-order interference.

For a non-entangled two-photon state (3.3.12), the first-order field correlation functions (3.3.15) become

$$\langle \psi_t|E_1^{(-)}(x_1)E_1^{(+)}(x_2)|\psi_t\rangle \propto C_1^*(x_1)C_1(x_2), \quad (3.3.17a)$$

$$\langle \psi_t|E_2^{(-)}(x_1)E_2^{(+)}(x_2)|\psi_t\rangle \propto C_2^*(x_1)C_2(x_2). \quad (3.3.17b)$$

Both modes satisfy the first-order complete coherence condition. As for an entangled two-photon state, each mode does not exhibit first-order coherence. In particular, for the perfect two-photon entanglement of Eq. (3.3.14), both modes are completely spatially incoherent:

$$\langle \psi_t|E_1^{(-)}(x_1)E_1^{(+)}(x_2)|\psi_t\rangle = \langle \psi_t|E_2^{(-)}(x_1)E_2^{(+)}(x_2)|\psi_t\rangle \propto \delta(x_1 - x_2). \quad (3.3.18)$$

As a matter of fact, each photon in a two-photon entangled state lies in a mixed state, such as Eq. (3.3.5).

For any two-photon state (3.3.9a), the second-order correlation function can be factorized into two two-photon amplitudes

$$\langle \psi_t|E_1^{(-)}(x_1)E_2^{(-)}(x_2)E_2^{(+)}(x_2')E_1^{(+)}(x_1')|\psi_t\rangle$$
$$= \langle \psi_t|E_1^{(-)}(x_1)E_2^{(-)}(x_2)|0\rangle \times \langle 0|E_2^{(+)}(x_2')E_1^{(+)}(x_1')|\psi_t\rangle, \quad (3.3.19a)$$

$$\langle \psi_t|E_1^{(-)}(x_1)E_2^{(-)}(x_2)E_2^{(+)}(x_2)E_1^{(+)}(x_1)|\psi_t\rangle = \left|\langle 0|E_2^{(+)}(x_2)E_1^{(+)}(x_1)|\psi_t\rangle\right|^2, \quad (3.3.19b)$$

where Eq. (3.3.19b) is the coincidence counting rate of two photons. For the non-entangled state (3.3.12), the above equation can be further factorized into four one-photon amplitudes

$$\langle \psi_t|E_1^{(-)}(x_1)E_2^{(-)}(x_2)E_2^{(+)}(x_2)E_1^{(+)}(x_1)|\psi_t\rangle = \langle \psi_{s1}|\langle \psi_{s2}|E_1^{(-)}(x_1)E_2^{(-)}(x_2)E_2^{(+)}(x_2)E_1^{(+)}(x_1)|\psi_{s2}\rangle|\psi_{s1}\rangle$$
$$= \langle \psi_{s1}|E_1^{(-)}(x_1)|0\rangle \times \langle \psi_{s2}|E_2^{(-)}(x_2)|0\rangle \times \langle 0|E_2^{(+)}(x_2)|\psi_{s2}\rangle \times \langle 0|E_1^{(+)}(x_1)|\psi_{s1}\rangle$$
$$= \left|\langle 0|E_1^{(+)}(x_1)|\psi_{s1}\rangle\right|^2 \times \left|\langle 0|E_2^{(+)}(x_2)|\psi_{s2}\rangle\right|^2. \quad (3.3.20)$$

Similar to a single-photon multimode mixed state of Eq. (3.3.5), a general two-photon mixed state is written with the density operator

$$\rho_t = \sum_{p_1,p_2} Q(p_1,p_2) a_1^\dagger(p_1) a_2^\dagger(p_2)|0\rangle\langle 0|a_2(p_2)a_1(p_1). \quad (3.3.21)$$

Here we consider the two-photon mixed state where two photons are indistinguishable when they are in the same spatial mode and all the sub-states have the same probability (a flat spatial spectrum)

$$\rho_{ti} = \sum_{p_1,p_2} a^\dagger(p_1) a^\dagger(p_2)|0\rangle\langle 0|a(p_2)a(p_1). \quad (3.3.22)$$

The first-order and second-order correlation functions can be calculated, respectively, from

$$\langle a^\dagger(p_1)a(p_2)\rangle = \mathrm{Tr}[\rho_{ti} a^\dagger(p_1)a(p_2)] \propto \delta(p_1 - p_2), \quad (3.3.23)$$

$$\langle a^\dagger(p_1)a^\dagger(p_2)a(p_2')a(p_1')\rangle = \mathrm{Tr}[\rho_{ti} a^\dagger(p_1)a^\dagger(p_2)a(p_2')a(p_1')]$$

$$= \langle a^\dagger(p_1)a(p_1')\rangle\langle a^\dagger(p_2)a(p_2')\rangle + \langle a^\dagger(p_1)a(p_2')\rangle\langle a^\dagger(p_2)a(p_1')\rangle. \tag{3.3.24}$$

Using Eq. (3.3.2), the corresponding first-order and second-order field correlation functions are, respectively, written as

$$\langle E^{(-)}(x_1)E^{(+)}(x_2)\rangle \propto \delta(x_1-x_2), \tag{3.3.25}$$

$$\langle E^{(-)}(x_1)E^{(-)}(x_2)E^{(+)}(x_2)E^{((+)}(x_1)\rangle$$
$$= \langle E^{(-)}(x_1)E^{(+)}(x_1)\rangle\langle E^{(-)}(x_2)E^{(+)}(x_2)\rangle + \left|\langle E^{(-)}(x_1)E^{(+)}(x_2)\rangle\right|^2. \tag{3.3.26}$$

The two-photon mixed state (3.3.22) satisfies the spatial incoherence condition (3.3.25) and the same intensity correlation (3.3.26) as the thermal light source (see Eq. (3.1.3)). This is mainly due to the fact that the two photons are in a flat spatial spectrum, an essential feature for chaotic light. For this reason, some authors regard the two-photon mixed state (3.3.22) as the quantum model to understand HB-T effect (Scarcelli et al., 2006).

Finally, we focus on biphoton coherence. A pair of entangled photons is sometimes called a biphoton (Klyshko, 1988). When Eq. (3.3.19a) is applied to a two-photon entangled state, the two-photon amplitude $\langle\psi_t|E_1^{(-)}(x_1)E_2^{(-)}(x_2)|0\rangle = C^*(x_1,x_2)$ cannot be further factorized. Regarding the two-photon amplitude as a whole, Eq. (3.3.19a) is similar to the expression of the first-order field coherence, and can be described as "*biphoton coherence*". Biphoton coherence is a kind of quantum coherence without classical correspondence. All the second-order interference effects of two-photon entangled state can be understood with biphoton coherence. Since the perfect entangled state of Eq. (3.3.14) displays complete incoherence for each field (see Eq. (3.3.18)), we consider the biphoton source as an incoherent source.

4. **Second-order interference with a two-photon source**

A typical second-order interference scheme for a two-photon state is similar to the configuration of Fano model. In the quantum description the field operators at the two detectors $D_1$ and $D_2$ are written, respectively, as

$$b_1(\omega_1) = a_1(\omega_1)\Phi_{11} + a_2(\omega_1)\Phi_{12}, \tag{3.4.1a}$$
$$b_2(\omega_2) = a_1(\omega_2)\Phi_{21} + a_2(\omega_2)\Phi_{22}, \tag{3.4.1b}$$

where $a_i$ is the field operator of source $S_i$ and $\Phi_{ji}$ is the optical transfer function from source $S_i$ to detector $D_j$, as shown in Fig. 3-2. Generally, two photons may have different frequencies and we assume that each detector records one particular frequency. The number operators at the detectors are

$$b_1^\dagger(\omega_1)b_1(\omega_1) = a_1^\dagger(\omega_1)a_1(\omega_1)|\Phi_{11}|^2 + a_2^\dagger(\omega_1)a_2(\omega_1)|\Phi_{12}|^2$$
$$+[a_1^\dagger(\omega_1)a_2(\omega_1)\Phi_{11}^*\Phi_{12} + \text{h.c.}], \tag{3.4.2a}$$
$$b_2^\dagger(\omega_2)b_2(\omega_2) = a_1^\dagger(\omega_2)a_1(\omega_2)|\Phi_{21}|^2 + a_2^\dagger(\omega_2)a_2(\omega_2)|\Phi_{22}|^2$$
$$+[a_1^\dagger(\omega_2)a_2(\omega_2)\Phi_{21}^*\Phi_{22} + \text{h.c.}]. \tag{3.4.2b}$$

We consider two types of two-photon entangled states emitted by the sources:

$$|\Psi_\text{I}\rangle = |\Psi_\text{I}^{(1)}\rangle + |\Psi_\text{I}^{(2)}\rangle = \tfrac{1}{\sqrt{2}}\left[a_1^\dagger(\omega_1)a_1^\dagger(\omega_2) + e^{i\beta}a_2^\dagger(\omega_1)a_2^\dagger(\omega_2)\right]|0\rangle, \tag{3.4.3a}$$
$$|\Psi_\text{II}\rangle = |\Psi_\text{II}^{(1)}\rangle + |\Psi_\text{II}^{(2)}\rangle = \tfrac{1}{\sqrt{2}}\left[a_1^\dagger(\omega_1)a_2^\dagger(\omega_2) + e^{i\beta}a_1^\dagger(\omega_2)a_2^\dagger(\omega_1)\right]|0\rangle. \tag{3.4.3b}$$

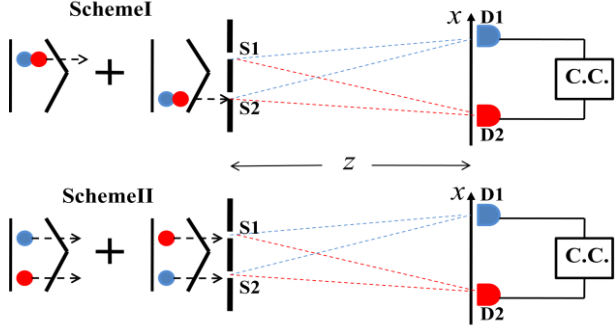

Fig. 3-3 Two types of two-photon entangled states partaking in second-order interference.

As shown in Fig. 3-3, $|\Psi_\text{I}\rangle$ describes both photons together at either of the sources in Scheme I, while $|\Psi_\text{II}\rangle$ denotes each photon at a source in Scheme II. These quantum states establish certain coherence between the two sources. Similar to Eq. (3.3.16), it is straight forward to verify that $\langle a_1^\dagger(\omega_i) a_2(\omega_i) \rangle = 0, (i=1,2)$ for both $|\Psi_\text{I}\rangle$ and $|\Psi_\text{II}\rangle$. As shown in Eq. (3.4.2), first-order interference does not show up in any single-photon detection. However, interference can occur in the second-order field correlation due to the biphoton coherence.

We can now calculate the two-photon coincidence counting rates for the two states. It is convenient to use the two-photon amplitude description. From Eq. (3.4.1) we have

$$b_1(\omega_1)b_2(\omega_2) = a_1(\omega_1)a_1(\omega_2)\Phi_{11}\Phi_{21} + a_2(\omega_1)a_2(\omega_2)\Phi_{12}\Phi_{22}$$
$$+ a_1(\omega_1)a_2(\omega_2)\Phi_{11}\Phi_{22} + a_2(\omega_1)a_1(\omega_2)\Phi_{12}\Phi_{21}. \tag{3.4.4}$$

The four terms in the right hand part of the above equation signify all the paths of the two-photon amplitudes: from $S_1$ to both $D_1$ and $D_2$, from $S_2$ to both $D_1$ and $D_2$, $S_1$ to $D_1$ while $S_2$ to $D_2$, and $S_1$ to $D_2$ while $S_2$ to $D_1$. Considering the two sub-states in both $|\Psi_\text{I}\rangle$ and $|\Psi_\text{II}\rangle$, we obtain, respectively, the two-photon amplitudes

$$\langle 0|b_1(\omega_1)b_2(\omega_2)|\Psi_\text{I}\rangle = \langle 0|a_1(\omega_1)a_1(\omega_2)|\Psi_\text{I}^{(1)}\rangle \Phi_{11}\Phi_{21} + \langle 0|a_2(\omega_1)a_2(\omega_2)|\Psi_\text{I}^{(2)}\rangle \Phi_{12}\Phi_{22}$$
$$= \tfrac{1}{\sqrt{2}}(\Phi_{11}\Phi_{21} + e^{i\beta}\Phi_{12}\Phi_{22}), \tag{3.4.5a}$$

$$\langle 0|b_1(\omega_1)b_2(\omega_2)|\Psi_\text{II}\rangle = \langle 0|a_1(\omega_1)a_2(\omega_2)|\Psi_\text{II}^{(1)}\rangle \Phi_{11}\Phi_{22} + \langle 0|a_2(\omega_1)a_1(\omega_2)|\Psi_\text{II}^{(2)}\rangle \Phi_{12}\Phi_{21}$$
$$= \tfrac{1}{\sqrt{2}}(\Phi_{11}\Phi_{22} + e^{i\beta}\Phi_{12}\Phi_{21}). \tag{3.4.5b}$$

Eventually, each sub-state in $|\Psi_\text{I}\rangle$ and $|\Psi_\text{II}\rangle$ possesses a corresponding two-photon path and the two-photon amplitude at the interference plane is the superposition of two sub-states. The two-photon coincidence counting rates for the two states are

$$\langle \Psi_\text{I}|b_1^\dagger(\omega_1)b_2^\dagger(\omega_2)b_1(\omega_1)b_2(\omega_2)|\Psi_\text{I}\rangle$$
$$= \tfrac{1}{2}[|\Phi_{11}\Phi_{21}|^2 + |\Phi_{12}\Phi_{22}|^2 + (\Phi_{11}^*\Phi_{21}^*\Phi_{12}\Phi_{22} e^{i\beta} + \text{c.c.})], \tag{3.4.6a}$$

$$\langle \Psi_\text{II}|b_1^\dagger(\omega_1)b_2^\dagger(\omega_2)b_1(\omega_1)b_2(\omega_2)|\Psi_\text{II}\rangle$$
$$= \tfrac{1}{2}[|\Phi_{11}\Phi_{22}|^2 + |\Phi_{12}\Phi_{21}|^2 + (\Phi_{11}^*\Phi_{22}^*\Phi_{12}\Phi_{21} e^{i\beta} + \text{c.c.})]. \tag{3.4.6b}$$

Thus quantum interference involves coherent superposition of two sub-states of a quantum state with the interference term of two biphoton amplitudes $\langle \Psi^{(1)}|b_1^\dagger(\omega_1)b_2^\dagger(\omega_2)|0\rangle \times \langle 0|b_1(\omega_1)b_2(\omega_2)|\Psi^{(2)}\rangle$, which includes a phase $\beta$ in the quantum states of Eq. (3.4.3).

Each individual sub-state in $|\Psi_I\rangle$ and $|\Psi_{II}\rangle$ is un-entangled two-photon state and will not exist the second-order interference except the degenerate case of $\omega_1=\omega_2$ for $|\Psi_{II}\rangle$, i.e.

$$|\Psi_{II}^{(\deg)}\rangle \equiv a_1^\dagger(\omega)a_2^\dagger(\omega)|0\rangle. \quad (3.4.7)$$

In this case two independent photons experiences two different paths, $\Phi_{11}\Phi_{22}$ and $\Phi_{12}\Phi_{21}$, simultaneously, so Eq. (3.4.6b) is reduced to

$$\langle\Psi_{II}^{(\deg)}|b_1^\dagger b_2^\dagger b_1 b_2|\Psi_{II}^{(\deg)}\rangle = |\Phi_{11}\Phi_{22}|^2 + |\Phi_{12}\Phi_{21}|^2 + (\Phi_{11}^*\Phi_{22}^*\Phi_{12}\Phi_{21} + \text{c.c.}). \quad (3.4.8)$$

In comparison of Eq. (3.4.8) to Eq. (3.4.6), they are very similar except a quantum phase $\beta$ in Eq. (3.4.6). This signifies the quantum interference nature for a two-photon entangled state, where the interference occurs between two quantum sub-states.

We now return to the original Fano model in which two source-fields with the same frequency are phase-independent. From Eq. (3.4.1) in the quantum description we obtain

$$\langle b_1^\dagger b_1 b_2^\dagger b_2\rangle = \langle a_1^\dagger a_1 a_1^\dagger a_1\rangle|\Phi_{11}\Phi_{21}|^2 + \langle a_2^\dagger a_2 a_2^\dagger a_2\rangle|\Phi_{12}\Phi_{22}|^2$$
$$+ \langle a_1^\dagger a_1 a_2^\dagger a_2\rangle\left(|\Phi_{11}\Phi_{22}|^2 + |\Phi_{12}\Phi_{21}|^2\right)$$
$$+ \left[\langle a_1^\dagger a_2 a_2^\dagger a_1\rangle\Phi_{11}^*\Phi_{22}^*\Phi_{12}\Phi_{21} + \langle a_2^\dagger a_1 a_1^\dagger a_2\rangle\Phi_{11}\Phi_{22}\Phi_{12}^*\Phi_{21}^*\right], \quad (3.4.9)$$

which is similar to Eq. (3.2.3) in the classical representation with the exception of shot noise. Applying it to the state composed of two independent photons $|\Psi_{II}^{(\deg)}\rangle$, we obtain the same result as Eq. (3.4.8). Therefore, two independent and undistinguishable photons follow Fano model in the second-order interference.

The above theoretical analysis tells us the physical difference between the entangled photon pair and the independent photon pair in the two-photon interference. For the former, the interference exists between two biphoton amplitudes of two sub-states. As for two independent photons, they comply with the original Fano model. In the following we shall discuss two important examples, beamsplitter interference and double-slit interference.

(i) *The Hong-Ou-Mandel interferometer*

In 1987, Hong, Ou and Mandel proposed a two-photon interference experiment through a beamsplitter (BS), later known as the Hong-Ou-Mandel (HOM) interferometer (Hong et al., 1987). As illustrated in Fig. 3-4(a), the signal and idler photons emitted from a UV laser-pumped crystal serving as a parametric down-converter are sent in opposite directions through a 50/50 BS. The emerging photon pair is allowed to impinge on two photon detectors $D_1$ and $D_2$, which perform the coincidence counting measurement. By moving BS continuously, the coincidence counts as a function of the BS displacement were registered. In the first HOM interferometer experiment the two photons to be interfered were degenerate with the same frequency and polarization. In Fig. 3-4(b) the experimental results show that for a certain symmetric position of BS, the two-photon coincidence rate falls to almost zero —this was later called the HOM dip (Hong et al., 1987). In further experiments it has been found that, in addition to the degenerate photons, interference may occur for two photons of different colour and polarization. For example, in the same experimental setup as in Fig. 3-4(a), the passbands of the two filters IF1 and IF2 may be centered on different frequencies $\omega_1$ and $\omega_2$ which do not overlap. As shown in Fig. 3-4(c), the two-photon coincidence count rate is found to exhibit a frequency beating of $\omega_1-\omega_2$ (Ou and Mandel, 1988).

It is necessary to define two types of two-photon interference that occur at a beamsplitter:

coalescence and anti-coalescence, according to whether the coincidence probabilities are less or more than that in the absence of interference, respectively. The experimental results for the degenerate two-photon state show coalescence interference, whereas for the non-degenerate two-photon entangled state, both coalescence and anti-coalescence interference may occur.

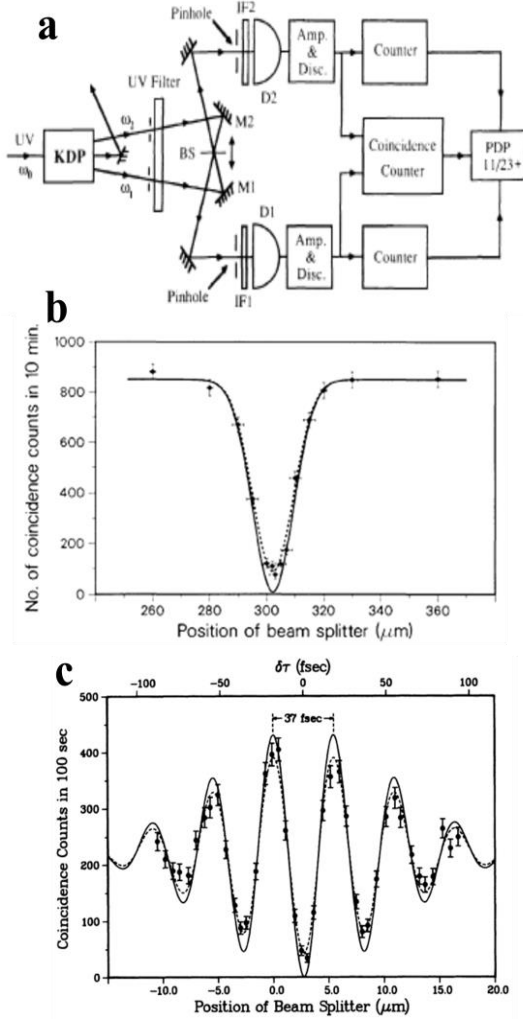

Fig. 3-4(a) Experimental setup of Hong-Ou-Mandel interferometer; (b) Experimental result for the degenerate two-photon state. The measured number of coincidences as a function of beam-splitter displacement (Hong et al., 1987). (c) Experimental result for the non-degenerate two-photon entangled state (Ou and Mandel, 1988).

In 2002, Santori et al. reported an HOM interferometer experiment in which two independent single-photon pulses were generated from a semiconductor quantum dot where photon entanglement was absent (Santori et al., 2002). Similarly, they observed a coalescence interference pattern with an HOM dip. Inspired by these experiments, later theoretical analysis demonstrated that anti-coalescence interference is the signature of two-photon entanglement (Wang, 2006). The HOM interferometer could be an effective experimental method to identify two-photon entanglement in addition to Bell's inequality test.

It is noticeable that in 2011 Chen et al. used two independent thermal light sources in the interferometer and observed a similar HOM dip (Chen et al., 2011). However, they considered the anticorrelation dip as quite a surprise and unexpected, and concluded that "the classical statistical correlation theory seems unable to explain this experimental result".

All these experiments can be interpreted with the above theoretical description. For a lossless beamsplitter, two input fields ($a_1$, $a_2$) and two output fields ($b_1$, $b_2$) satisfy Eq. (3.4.1). The transfer functions of the BS are given by

$$\Phi_{11} = e^{i\varphi_\tau}\cos\theta, \quad \Phi_{12} = e^{i\varphi_\rho}\sin\theta, \quad \Phi_{21} = -e^{-i\varphi_\rho}\sin\theta, \quad \Phi_{22} = e^{-i\varphi_\tau}\cos\theta, \quad (3.4.10)$$

where $\cos^2\theta$ ($\sin^2\theta$) stands for the transmissivity (reflectivity) of BS, and $\varphi_\tau$ and $\varphi_\rho$ are its two phase parameters. Note that Eq. (3.4.10) follows the unitary transformation, so the two output fields can commutate each other. Using Eqs. (3.4.2) and (3.4.10) we obtain two output intensities

$$b_1^\dagger b_1 = a_1^\dagger a_1 \cos^2\theta + a_2^\dagger a_2 \sin^2\theta + [a_1^\dagger a_2 e^{i\varphi} + \text{h.c.}]\cos\theta\sin\theta, \quad (3.4.11a)$$

$$b_2^\dagger b_2 = a_1^\dagger a_1 \sin^2\theta + a_2^\dagger a_2 \cos^2\theta - [a_1^\dagger a_2 e^{i\varphi} + \text{h.c.}]\cos\theta\sin\theta, \quad (3.4.11b)$$

where $\varphi = \varphi_\rho - \varphi_\tau$. When the two input fields are coherent with each other, $\langle a_1^\dagger a_2 \rangle \neq 0$, then the intensities of the two output fields are related to $\langle a_1^\dagger a_2 \rangle$ and first-order interference occurs. However, we have seen from above that for two independent fields or a two-photon state $|\Psi_{\text{II}}\rangle$, first-order interference does not occur at a BS, and the intensities of the two output fields depend on the transmissivity and reflectivity only. In this sense the HOM interferometer should be included in the class of incoherent interference.

Using Eq. (3.4.11) in Eq. (3.4.6b) and Eq. (3.4.8), we obtain

$$\langle\Psi_{\text{II}}|b_1^\dagger(\omega_1)b_2^\dagger(\omega_2)b_1(\omega_1)b_2(\omega_2)|\Psi_{\text{II}}\rangle = \tfrac{1}{2}(\cos^4\theta + \sin^4\theta - 2\cos^2\theta\sin^2\theta\cos\beta) \quad (3.4.12)$$

and

$$\langle\Psi_{\text{II}}^{(\text{deg})}|b_1^\dagger b_2^\dagger b_1 b_2|\Psi_{\text{II}}^{(\text{deg})}\rangle = (\cos^2\theta - \sin^2\theta)^2, \quad (3.4.13)$$

respectively. In the degenerate case the coincidence count rate reaches zero for a 50/50 BS. However, in the non-degenerate case, both coalescence and anti-coalescence interference can occur depending on the phase $\beta$. When the two input paths have different lengths, $l_1$ and $l_2$, $\beta = (\omega_1 - \omega_2)(l_2 - l_1)$ leads to frequency beating.

For the HOM interference of two independent thermal light sources, using Eqs. (3.4.9) and (3.4.10), we obtain the intensity fluctuation correlation of two output fields to be

$$\langle b_1^\dagger b_1 b_2^\dagger b_2\rangle - \langle b_1^\dagger b_1\rangle\langle b_2^\dagger b_2\rangle = |\langle b_1^\dagger b_2\rangle|^2 = (\langle a_1^\dagger a_1\rangle - \langle a_2^\dagger a_2\rangle)^2 \cos^2\theta\sin^2\theta. \quad (3.4.14)$$

As for two independent coherent sources, we have

$$\langle b_1^\dagger b_1 b_2^\dagger b_2\rangle - \langle b_1^\dagger b_1\rangle\langle b_2^\dagger b_2\rangle = |\langle b_1^\dagger b_2\rangle|^2 = -2\langle a_1^\dagger a_1\rangle\langle a_2^\dagger a_2\rangle\cos^2\theta\sin^2\theta. \quad (3.4.15)$$

Equation (3.4.14) shows that when the two input thermal light beams have equal intensity, the two output thermal light beams are no longer correlated. This equation presents a better understanding for the experimental result in 2011(Chen et al., 2011). It is interesting that the HOM interferometer for two independent coherent sources reveals negative correlation. Therefore we can say that for two independent sources this interferometer always decreases the correlation of the original sources. However, this conclusion cannot apply to a two-photon entangled state since the photons in an entangled state are not independent.

(ii) *Second-order double-slit interference experiments*

The first experiment on the interference of light was Young's double-slit setup, which demonstrated the wave nature of light. As a second example of second-order interference, we return to the double-slit experiment. The first tests on Young's double-slit interference with a two-photon entangled state were reported around 2000 (Fonseca et al., 1999a; D'Angelo et al., 2001). The subwavelength characteristic of the second-order interference fringes attracted wide attention. In 2005 Xiong et al. reported observing subwavelength interference with a pseudo-thermal light source (Xiong et al., 2005), which was also observed with a true thermal light source (Zhai et al., 2005).

Later, a second-order double-slit experiment with two-color biphotons was reported (Zhang et al., 2017). In classical optics, Young's double-slit experiment with colored coherent light gives rise to individual interference fringes for each frequency. When two colored photons are entangled, double-slit interference fringes cannot be observed for either color individually. Instead however, interference information can be extracted through the coincidence measurements of two colored photons (Zhang et al., 2017).

Different from the single-photon case, for entangled biphoton pairs there are two possible interference configurations. As shown in Fig. 3-3, both photons pass together through either of the slits in Scheme I, and each photon passes through a different slit in Scheme II. The two schemes involve two kinds of biphoton entangled states at the double-slit, which have been already given in Eqs. (3.4.3a) and (3.4.3b). However, the previous two-photon double-slit experiments (Fonseca et al., 1999a; D'Angelo et al., 2001) involved Scheme I with two degenerate photons. For the non-degenerate case, the two detectors D1 and D2 in the observation plane are covered by filters F1 and F2 to select the frequencies $\omega_1$ and $\omega_2$, respectively. Although each detector can identify the photon color, it does not know which slit the photon came from. As a result, the fields recorded by the detectors D1 and D2 follow Eq. (3.4.1), with

$$\Phi_{11} = \exp(i\omega_1 r_{11}/c), \quad \Phi_{12} = \exp(i\omega_1 r_{12}/c),$$
$$\Phi_{21} = \exp(i\omega_2 r_{21}/c), \quad \Phi_{22} = \exp(i\omega_2 r_{22}/c), \qquad (3.4.16)$$

where $r_{ij}$ (i,j=1,2) is the distance between detector Di and slit Sj, and $c$ is the speed of light.

As pointed out at the beginning of the present subsection, the first-order correlation functions for both $|\Psi_I\rangle$ and $|\Psi_{II}\rangle$ vanish. The same as in the HOM interferometer, all the one-photon interference effects do not exist, i.e. $\langle\Psi_j|b_i^\dagger(\omega_i)b_i(\omega_i)|\Psi_j\rangle =$ constant for i=1,2 and j=I,II. Hence, for both schemes, an individual detector Di does not detect interference fringes. In fact, as shown in Eq. (3.3.18), for a perfect two-photon entangled state each photon displays spatial incoherence.

Instead, the two-photon amplitudes in the two schemes do display the spatial coherence. Substituting Eq. (3.4.16) into Eq. (3.4.5), we obtain

$$\langle 0|b_1(\omega_1)b_2(\omega_2)|\Psi_I\rangle$$
$$= \tfrac{\sqrt{2}}{2}\{\exp[i(\omega_1 r_{11}+\omega_2 r_{21})/c] + \exp[i(\omega_1 r_{12}+\omega_2 r_{22})/c+i\beta]\}, \qquad (3.4.17a)$$
$$\langle 0|b_1(\omega_1)b_2(\omega_2)|\Psi_{II}\rangle$$
$$= \tfrac{\sqrt{2}}{2}\{\exp[i(\omega_1 r_{11}+\omega_2 r_{22})/c] + \exp[i(\omega_1 r_{12}+\omega_2 r_{21})/c+i\beta]\}. \qquad (3.4.17b)$$

Obviously, in the two-photon amplitudes of Eqs. (3.4.17a) and (3.4.17b), there are two possible paths, which originate from the two sub-states of the two-photon entangled states. The coherent superposition of the two two-photon amplitudes gives rise to the second-order interference. The coincidence count rates at the two detectors for the two schemes are readily obtained from:

$$\langle\Psi_I|b_1^\dagger(\omega_1)b_2^\dagger(\omega_2)b_1(\omega_1)b_2(\omega_2)|\Psi_I\rangle = 1+\cos[(\omega_1 x_1+\omega_2 x_2)d/(cz)]$$
$$= 1+\cos\{[\omega_+(x_1+x_2)+\omega_-(x_1-x_2)]d/(2cz)\}, \qquad (3.4.18a)$$
$$\langle\Psi_{II}|b_1^\dagger(\omega_1)b_2^\dagger(\omega_2)b_1(\omega_1)b_2(\omega_2)|\Psi_{II}\rangle = 1+\cos[(\omega_1 x_1-\omega_2 x_2)d/(cz)]$$
$$= 1+\cos\{[\omega_+(x_1-x_2)+\omega_-(x_1+x_2)]d/(2cz)\}, \qquad (3.4.18b)$$

where $d$ is the slit spacing, $z$ the distance between the double-slit and the observation plane, $x_i$ the position of detector Di in the observation plane, $\omega_+ = \omega_1+\omega_2$ is the sum-frequency, and $\omega_- = \omega_1-\omega_2$ is the difference-frequency. These results are quite different from those of conventional first-order double-slit interference.

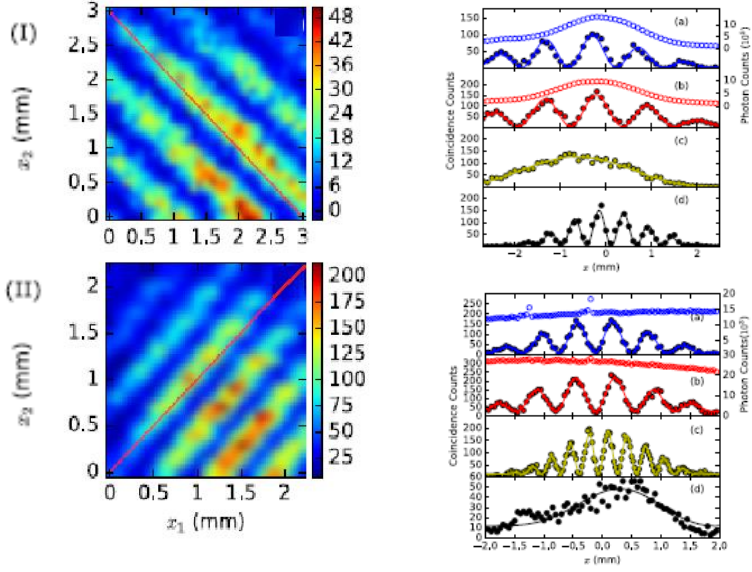

Fig. 3-5 Experimental results of two-color biphoton double-slit interference (wavelengths 760 nm and 840 nm). Above for Scheme I and below for Scheme II. Left: the two-axis ($x_1$ by $x_2$) coincidence counting interference patterns. Right: four detection outcomes, where open-circles are single-photon counts and full-circles are coincidence counts, where (a)D1 is scanned while D2 is fixed; (b)D2 is scanned while D1 is fixed; (c)D1 and D2 are scanned oppositely; (d) D1 and D2 are scanned together (Zhang et al., 2017).

Figure 3-5 shows the experimental results of the two-color double-slit interference (Zhang et al., 2017). The entire information of the second-order interference can be revealed in a two-axis ($x_1$ by $x_2$) coincidence counting interference pattern, as shown in the left part of Fig. 3-5. The fringe stripes for Scheme I (above) and Scheme II (below) are basically along the anti-diagonal and diagonal directions, respectively. A slight deviation from the diagonal lines is due to the fact that the two photons have different colors, 760nm and 840nm. In fact, for the frequency degenerate case, the fringe stripes for both Schemes I and II are exactly along the anti-diagonal and diagonal directions, respectively (see Ref. [(Zhang et al., 2017)]).

In the two-axis interference pattern there are four possible measurement outcomes, which manifest the interference fringes at four frequencies: $\omega_1$ and $\omega_2$, the sum-frequency $\omega_+$ and the difference-frequency $\omega_-$. When detector D2 (D1) is fixed and D1 (D2) is scanned, the interference pattern for frequency $\omega_1$ ($\omega_2$) is represented by the blue (red) lines for both schemes. However, when both detectors are moved together in the same direction (scanning along the diagonal line), one should see the interference fringes (black lines) of the sum-frequency $\omega_+$ for Scheme I and the difference-frequency $\omega_-$ for Scheme II. On the contrary, when the two detectors are moved in opposite directions (scanning along the anti-diagonal line), one see the interference fringes (yellow lines) of difference-frequency $\omega_-$ for Scheme I and sum-frequency $\omega_+$ for Scheme II. These results can be readily understood from Eqs. (3.4.18a) and (3.4.18b).

We can now understand why the subwavelength interference for Scheme I observed in previous studies (Fonseca et al., 1999a; D'Angelo et al., 2001) is actually the result of sum-frequency interference for degenerate photon pairs while the difference-frequency effect disappears.

5. **Second-order interference with thermal light**

At the beginning of this century, both theory and experiment demonstrated that thermal light can mimic two-photon entangled states in performing ghost imaging, ghost interference and subwavelength interference (Bennink et al., 2002; Gatti et al., 2004; Cheng and Han, 2004; Wang and Cao, 2004; Cai and Zhu, 2004; Bennink et al., 2004; Cao et al., 2005; Xiong et al., 2005; Valencia et al., 2005; Ferri et al., 2005; Zhai et al., 2005). These effects, along with the original HB-T experiment, can be explained by the intensity correlation equation (3.1.3).

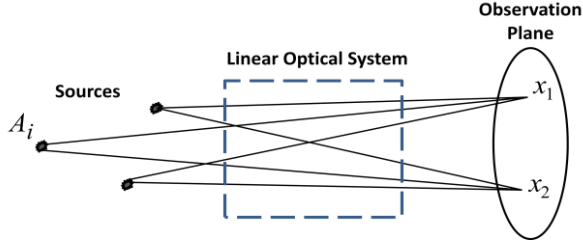

Fig. 3-6 Superposed field by statistically independent point sources.

To better understand the underlying physics, we consider an optical field, illuminated by $N$ statistically independent point light sources with the same wavelength and polarization, as shown in Fig. 3-6. The superposed field in the observation plane is $E(x) = \sum E_i(x)$, where $E_i(x)$ is the sub-field in the same plane, generated from the $i$-th point-source. Taking into account the statistical independence between the point sources, the first-order field correlation function at the two positions $x_1$ and $x_2$ in the observation plane is given by

$$\langle E^*(x_1)E(x_2)\rangle = \sum_i \langle E_i^*(x_1)E_i(x_2)\rangle. \tag{3.5.1}$$

If each point source satisfies temporal coherence, it contributes the first-order field correlation function to the superposed field. For $x_1=x_2=x$, Eq. (3.5.1) shows that first-order interference does not exist between the point sources, as it is usually referred to spatial incoherence of independent sources.

The intensity correlation at the two positions $x_1$ and $x_2$ in the observation plane is

$$\langle E^*(x_1)E^*(x_2)E(x_2)E(x_1)\rangle = \sum_{i,j,k,l}\langle E_i^*(x_1)E_j^*(x_2)E_k(x_2)E_l(x_1)\rangle$$

$$= \sum_{i,j(i\neq j)}\langle E_i^*(x_1)E_i(x_1)\rangle\langle E_j^*(x_2)E_j(x_2)\rangle + \sum_{i,j(i\neq j)}\langle E_i^*(x_1)E_i(x_2)\rangle\langle E_j^*(x_2)E_j(x_1)\rangle$$

$$+ \sum_i \langle E_i^*(x_1)E_i^*(x_2)E_i(x_2)E_i(x_1)\rangle$$

$$= \langle E^*(x_1)E(x_1)\rangle\langle E^*(x_2)E(x_2)\rangle + \left|\langle E^*(x_1)E(x_2)\rangle\right|^2$$

$$+ \sum_i \left\{\langle E_i^*(x_1)E_i^*(x_2)E_i(x_2)E_i(x_1)\rangle - \langle E_i^*(x_1)E_i(x_1)\rangle\langle E_i^*(x_2)E_i(x_2)\rangle - \left|\langle E_i^*(x_1)E_i(x_2)\rangle\right|^2\right\}. \tag{3.5.2}$$

For a linear optical system, the correlation functions in the braces of the last term are related to the intensity fluctuations of the point source as follows:

$$\langle E_i^*(x_1)E_i^*(x_2)E_i(x_2)E_i(x_1)\rangle - \langle E_i^*(x_1)E_i(x_1)\rangle\langle E_i^*(x_2)E_i(x_2)\rangle - \left|\langle E_i^*(x_1)E_i(x_2)\rangle\right|^2$$

$$\propto \langle |A_i|^4\rangle - 2\langle |A_i|^2\rangle, \tag{3.5.3}$$

where $A_i$ is the amplitude of the $i$-th source. If the intensity statistical distribution of all the point sources follows the negative exponential function for thermal source, the above equation is null, and Eq. (3.5.2) reduces to Eq. (3.1.3).

The above derivation reveals the nature of the intensity correlation of thermal light. The thermal light source is composed of an incoherent superposition of many independent point-sources. In free propagation each point-source generates a homogeneous intensity distribution while maintaining spatial coherence of any two given positions as long as their temporal coherence condition is fulfilled. The spatial coherence of the two positions in the superposed field disappears in the intensity observation but can be extracted through intensity correlation measurements.

In 2003, Cao and Wang found that the subwavelength Young's double-slit interference can also appear in a high-gain spontaneous parametric down conversion (SPDC) process with an opposite spatial correlation with respect to the low-gain case. They immediately realized that the spatial correlation in the high-gain case originates from the thermal emission statistics (Cao and Wang, 2004; Wang and Cao, 2004). Soon Xiong et al. performed an experiment on the second-order double-slit interference with a pseudo-thermal light source (Xiong et al., 2005; Wang et al., 2006).

To show the relation between the first-order spatial coherence and the second-order spatial correlation in the double-slit interference, we assume that both the double-slit and the observation screen are placed at the two focal planes of a lens of focal length $f$. Assume that a thermal light source illuminating a double-slit has a spatial spectral correlation $\langle E^*(q)E(q')\rangle = S(q)\delta(q-q')$ with an angular spectrum $S(q) = (\sqrt{2\pi}w)^{-1}\exp[-q^2/(2w^2)]$, where $q$ is the transverse wavevector and $w$ is the spatial spectral width. The first-order and second-order correlation functions in the detection plane are given by

$$G^{(1)}(x_1,x_2) = \langle E^*(x_1)E(x_2)\rangle = \frac{k}{2\pi f}\int \tilde{T}^*\left(\frac{kx_1}{f}-q\right)\tilde{T}\left(\frac{kx_2}{f}-q\right)S(q)\mathrm{d}q, \quad (3.5.4a)$$

$$G^{(2)}(x_1,x_2) = \langle E^*(x_1)E^*(x_2)E(x_2)E(x_1)\rangle = G^{(1)}(x_1,x_1)G^{(1)}(x_2,x_2) + |G^{(1)}(x_1,x_2)|^2, \quad (3.5.4b)$$

where $k$ is the wave number. $\tilde{T}(q) = (2b/\sqrt{2\pi})\mathrm{sinc}(qb/2)\cos(qd/2)$ is the Fourier transform of the double-slit function $T(x)$, and $d$ and $b$ are the slit spacing and slit width, respectively.

When the source has perfect spatial coherence with a plane wavefront, $S(q)=\delta(q)$, we can obtain the coherent interference patterns $G^{(1)}(x,x) \propto |\tilde{T}(kx/f)|^2$ and $G^{(2)}(x_1,x_2) \propto |\tilde{T}(kx_1/f)\tilde{T}(kx_2/f)|^2$. On the contrary, when the source has total incoherence, $S(q)=S_0$, we obtain

$$G^{(1)}(x_1,x_2) = \frac{k}{2\pi f}\sqrt{2\pi}S_0\tilde{T}\left[\frac{k}{f}(x_2-x_1)\right], \quad (3.5.5a)$$

$$G^{(2)}(x_1,x_2) = \left(\frac{k}{2\pi f}\right)^2 2\pi S_0^2\left\{\tilde{T}^2(0) + \left|\tilde{T}\left[\frac{k}{f}(x_2-x_1)\right]\right|^2\right\}. \quad (3.5.5b)$$

As a result, the first-order interference disappears, i.e. $G^{(1)}(x,x) \propto \tilde{T}(0)$. The interference pattern can be seen through the intensity correlation measurement at the two positions. When $x_2=-x_1=x$, the second-order interference pattern exhibits the subwavelength feature, $G^{(2)}(x,-x) \propto \tilde{T}^2(0) + |\tilde{T}(\frac{k}{f}2\mathrm{x})|^2$.

Figure 3-7 shows the normalized first-order and second-order interference patterns when the normalized spatial bandwidth $W=wd/(2\pi)$ is varied. For the coherent case with smaller bandwidth, the first-order and second-order interference patterns have no significant difference. As the spatial bandwidth increases, the first-order interference pattern disappears while the second-order interference pattern appears with a period half that of the first-order fringes. We see that the decrease of coherence is associated with the increase of the second-order correlation.

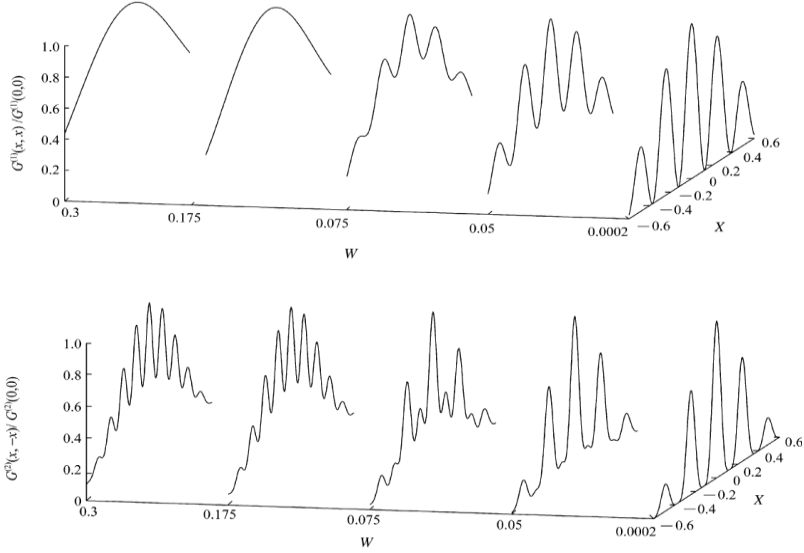

Fig. 3-7 The first-order (above) and second-order (below) double-slit interference patterns when the normalized spatial spectrum width $W$ is varied (Wang et al., 2006).

6. **Ghost diffraction and ghost imaging**

The most fascinating effects of two-photon entanglement light and thermal light are ghost diffraction and ghost imaging. Ghost diffraction and ghost imaging with a two-photon entangled source were theoretically proposed by Belinskii and Klyshko in 1994 (Belinskii and Klyshko, 1994), and then first experiments were implemented by Pittman et al. and Strekalov et al. in 1995 (Pittman et al., 1995; Strekalov et al., 1995). Later, it was demonstrated, both theoretically and experimentally, that ghost diffraction and ghost imaging can also be realized with thermal light (Bennink et al., 2002; Gatti et al., 2004; Cheng and Han, 2004; Wang and Cao, 2004; Cai and Zhu, 2004; Bennink et al., 2004; Cao et al., 2005b; Valencia et al., 2005; Ferri et al., 2005; Zhai et al., 2005).

The sketch of ghost diffraction is shown in Fig. 3-8, where the light from source S is divided into two parts by a beamsplitter BS. One beam (or photon) travels in the test arm, where an object T of transmittance $T(x)$ is located at a distance $z_{o1}$ from source S and $z_{o2}$ from detector $D_1$. The impulse response function of the test arm is written as

$$h_o(x, x_0) = \int H(x, x'; z_{o2}, z_{o2})T(x')H(x', x_0; z_{o1}, z_{o1})\mathrm{d}x', \qquad (3.6.1)$$

where function $H$ is defined in Eq. (1.3.3). The other beam (or photon) propagates freely in the reference arm of length $z_r$ with the impulse response function $h_r(x,x_0)=H(x,x_0;z_r,z_r)$.

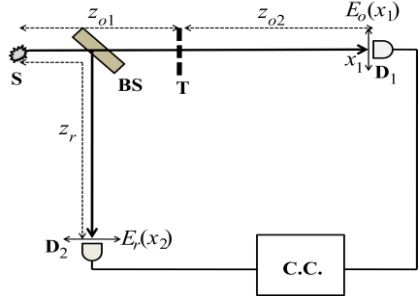

Fig. 3-8 Sketch of ghost diffraction: S, source; BS, beamsplitter; T, object; $D_1$ and $D_2$, two detectors; C.C., coincidence counting measurement.

If the source is in a two-photon entangled state, in which the two photons of mode 1 and mode 2 pass through the test arm and the reference arm, respectively, the two-photon amplitude at the two detection planes is written as

$$\langle 0|E_o^{(+)}(x_1)E_r^{(+)}(x_2)|\psi_t\rangle = \int h_o(x_1,x_0)h_r(x_2,x_0')\langle 0|E_1^{(+)}(x_0)E_2^{(+)}(x_0')|\psi_t\rangle \mathrm{d}x_0\mathrm{d}x_0' , \quad (3.6.2)$$

where $\langle 0|E_1^{(+)}(x_0)E_2^{(+)}(x_0')|\psi_t\rangle$ is the two-photon amplitude at the source. For the perfect entanglement, $\langle 0|E_1^{(+)}(x_0)E_2^{(+)}(x_0')|\psi_t\rangle \propto \delta(x_0-x_0')$ (see Eq. (3.3.14)), Eq. (3.6.2) becomes

$$\langle 0|E_o^{(+)}(x_1)E_r^{(+)}(x_2)|\psi_t\rangle = \int h_o(x_1,x_0)h_r(x_2,x_0)\mathrm{d}x_0$$
$$= \int H(x_1,x';z_{o2},z_{o2})T(x')H(x',x_2;z_{o1}+z_r,z_{o1}+z_r)\mathrm{d}x' . \quad (3.6.3)$$

The expression signifies *joint-field diffraction* (JFD) of two-photon amplitude in the perfect entanglement. By comparing the above equation with Eq. (3.6.1), it can be equivalent to the single photon case in which a point source at detector $D_1$ shines the object T and the diffraction pattern is observed at detector $D_2$, or vice versa. The coincidence counting measurement of two photons is proportional to

$$\left|\langle 0|E_o^{(+)}(x_1)E_r^{(+)}(x_2)|\psi_t\rangle\right|^2 \propto \left|\int T(x')\exp\left[\frac{ik_0}{2Z_{en}}\left(x'-\frac{(z_{o1}+z_r)x_1+z_{o2}x_2}{z_{o1}+z_{o2}+z_r}\right)^2\right]\mathrm{d}x'\right|^2 , \quad (3.6.4a)$$

where the effective diffraction length $Z_{en}$ is given by

$$\frac{1}{Z_{en}} = \frac{1}{z_{o2}} + \frac{1}{z_{o1}+z_r} . \quad (3.6.4b)$$

As for a thermal light source, the intensity fluctuation correlation at two detectors records the first-order field correlation function [see Eq. (1.3.2)]

$$\langle E_r^*(x_2)E_o(x_1)\rangle = \int h_r^*(x_2,x_0)h_o(x_1,x_0)\mathrm{d}x_0 . \quad (3.6.5a)$$

The expression is defined as *correlated-field diffraction* (CFD). For the scheme of Fig. 3-8, Eq. (3.6.5a) is calculated to be

$$\langle E_r^*(x_2)E_o(x_1)\rangle = \frac{I_0 k_0}{2\pi}\sqrt{\frac{k_0}{i2\pi z_{o1}z_{o2}z_r}}\exp[ik_0(z_{o1}+z_{o2}-z_r)]$$
$$\times \int T(x')\exp\left\{\frac{ik_0}{2}\left[\frac{(x_1-x')^2}{z_{o2}}+\frac{(x'-x_0)^2}{z_{o1}}-\frac{(x_2-x_0)^2}{z_r}\right]\right\}\mathrm{d}x_0\mathrm{d}x' . \quad (3.6.5b)$$

The general result of intensity correlation can be further written as

$$\left|\langle E_r^*(x_2)E_o(x_1)\rangle\right|^2 \propto \left|\int T(x')\exp\left[\frac{ik_0}{2Z_{th}}\left(x' - \frac{(z_{o1}-z_r)x_1 + z_{o2}x_2}{z_{o1}+z_{o2}-z_r}\right)^2\right]dx'\right|^2, \quad (3.6.6a)$$

where the effective diffraction length $Z_{th}$ is given by

$$\frac{1}{Z_{th}} = \frac{1}{z_{o2}} + \frac{1}{z_{o1}-z_r}. \quad (3.6.6b)$$

Both Eqs. (3.6.4) and (3.6.6) behave Fresnel diffraction for coherent light. The diffraction pattern can be observed with scanning either one of two detectors while the other is fixed.

For thermal light correlation, the effective diffraction length $Z_{th}$ can be negative for $z_{o1}<z_r<z_{o1}+z_{o2}$, designating a wavefront-reversal diffraction. However, the normal diffraction appears in other two regions, $z_r<z_{o1}$ and $z_r>z_{o1}+z_{o2}$. Very interesting phenomena appear at the two borders of diffraction transition. At $z_r=z_{o1}+z_{o2}$, the correlated-field diffraction yields exactly lensless Fourier transform (Gatti et al., 2004; Cheng and Han, 2004)

$$\left|\langle E_r^*(x_2)E_o(x_1)\rangle\right|^2 \propto \left|\int T(x')\exp\left[\frac{ik_0}{2z_r}(x_2-x_1)x'\right]dx'\right|^2. \quad (3.6.7)$$

In the special case of $z_r=z_{o1}$, however, the detector $D_2$ will record equal image of an object without using a lens, $\left|\langle E_r^*(x_2)E_o(x_1)\rangle\right|^2 \propto |T(x_2)|^2$. Lensless imaging is the distinction of phase conjugation effect that exists in the nonlinear optical process.

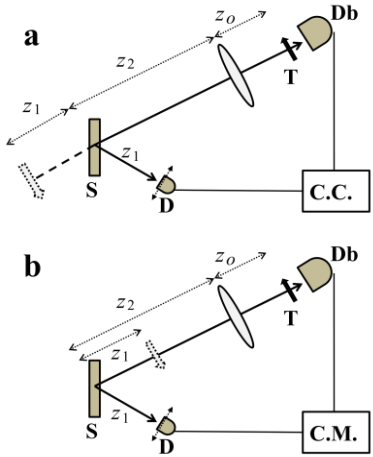

Fig. 3-9 Sketches of ghost imaging: S, source; T, object; Db, bucket detector; D, scanning detector; C.C., coincidence counting; C.M., correlation measurement. (a) S is a two-photon entangled source, which acts as a mirror; (b) S is a thermal light source, which acts as a phase-conjugation mirror.

Imaging with lens is the special diffraction of point-to-point correspondence. Figures 3-9(a) and 3-9(b) show ghost imaging schemes for a two-photon entangled source and a thermal light source, respectively. Due to joint-field diffraction and correlated-field diffraction, they follow, respectively, the lens-imaging equations

$$\frac{1}{z_o} + \frac{1}{z_2+z_1} = \frac{1}{f}, \quad (3.6.8a)$$

$$\frac{1}{z_o} + \frac{1}{z_2-z_1} = \frac{1}{f}, \quad (3.6.8b)$$

where $f$ is the focal length of the imaging lens, and $z_o$ is the distance from object to lens. $z_1$ and $z_2$ are the distances from source to scanning detector D and lens, respectively. With respect to the physical picture of joint-field diffraction and correlated-field diffraction, Cao et al. first pointed out that "The quantum entangled source behaves as a mirror, whereas the classical thermal source looks like a phase conjugate mirror in the correlated imaging." (Cao et al., 2005b).

7. **Summary of second-order interference**

Since second-order coherence is satisfied for a coherent optical field, both first-order interference and second-order interference coexist. However the latter does not provide more information than the former. In the examples of second-order interference discussed above, we used three sources: a pair of independent sources, a quantum entangled two-photon source, and a classical thermal light source. All of them display first-order spatial incoherence and hence first-order interference does not occur in the conventional interference configuration. In this sense, therefore, these examples of second-order interference, both classical and quantum, can be regarded as incoherent interference.

For these sources the underlying second-order interference mechanisms are not the same. Basically, a quantum entangled source can give rise to quantum interference where biphoton coherence plays a central role, while a classical thermal light source generates classical interference which is governed by the classical field correlation. A two-photon entangled state consists of a coherent superposition of a pair of two-photon sub-states. When a biphoton passes through an optical system, second-order interference occurs between the two sub-states (see Fig. 3-3). The concept of two-photon amplitude is directly related to biphoton coherence, which is a form of quantum coherence without classical correspondence. It should be noted that a pure state of two independent photons and a two-photon mixed state of wide spectral bandwidth both exhibit the same classical field correlation of two positions.

It is noteworthy that in the classical regime, such as when thermal light is used, each independent point source generates its own subsequent first-order spatial coherence, and coherent information is included in the first-order field correlation function of two positions (see Fig. 3-6). Accordingly, the interference has to be observed through the intensity correlation measurement at these two positions. In this sense we can say that in second-order interference for thermal light each photon interferes with itself while in quantum second-order interference for a two-photon entangled state, each biphoton interferes with itself.

IV. **First-order incoherent interference**

We now back to first-order interference. In Gabor's coherent holography, reference light is a part of the coherent illumination undisturbed by the object. Both amplitude and phase information of an object can be recorded in the interference pattern. However, incoherent holography discussed in Sec. II is different. The two fields joining the interference come from the same point-source of a fluorescent object. Only intensity distribution of the object can be recorded, as shown in Eq. (2.4). A challenging question would be: can Gabor's coherent holography be also performed using incoherent light? The following experiments give positive answer to the question.

1. **Incoherent interference in an unequal-optical-path interferometer**

In 2009, Zhang et al. reported an experiment in which incoherent illumination can be applied to similar holographic interference as Gabor's original conception (Zhang et al., 2009a). As pointed out in Sec. I-3 that when both object and reference beams come from a spatially incoherent source and then travel same distance, the information of object will be washed out completely in the holographic interference (see Eq. (1.3.4)). As a proof-of-principle experiment, they first use an unequal-optical-path interferometer driven by a spatial incoherent illumination. As shown in Fig. 4-1, a pseudo-thermal light source is formed by passing a He-Ne laser beam of wavelength 632.8 nm through a rotating ground glass disk G. The interferometer is designed to have unequal arm lengths: the object arm $z_o$=16cm and the reference arm $z_r$=27cm. The object in the experiment is a double-slit T of slit width $b$=125μm and spacing $d$=310μm. The interference pattern can be recorded by either of two charge-coupled device (CCD) cameras.

The laser beam has good temporal coherence which will not be affected by the scattering of the slowly rotating ground glass. The optical path difference between the two arms is much smaller than the longitudinal coherence length $c\tau$, where $c$ is the light speed in vacuum and $\tau$ is the coherence time of the laser beam. However, the motion of the ground glass disk destroys the spatial coherence of the laser beam. The source emits what is known as pseudo-thermal light and satisfies Eq. (1.3.1) approximately. In contrast, a true thermal light source with coherence time less than $10^{-10}$ sec. cannot satisfy the condition of temporal coherence in this setup, and the interference will not occur.

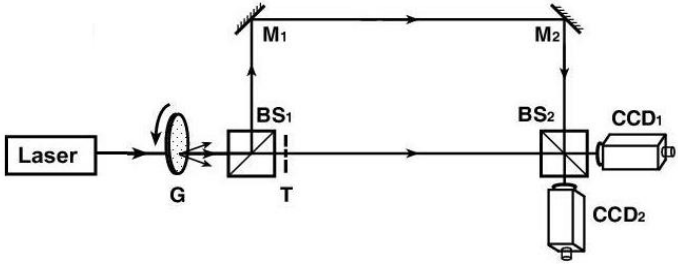

Fig. 4-1 Experimental scheme for an unequal-optical-path interferometer where the two arms have different lengths. A laser and a rotating ground glass disk G form a spatially incoherent source; $M_1$ and $M_2$ are mirrors, $BS_1$ and $BS_2$ are beamsplitters, $CCD_1$ and $CCD_2$ are detectors, and T is a double-slit close to $BS_1$ (Zhang et al., 2009a).

The two-dimensional (2D) intensity patterns detected by $CCD_1$ are shown in Fig. 4-2. When the ground glass disk is moved step by step, the interference patterns are irregular speckles frame by frame and each frame is different, as shown in the two single-shot frames of Figs. 4-2(a) and 4-2(b). However, if a number of exposures are superposed, a well-defined interference pattern emerges gradually in Figs. 4-2(c)–4-2(g) as the number of frames increases.

The experimental results tell us the basic physical feature of incoherent interference: the interference pattern is irregularly fluctuating in time and the well-defined pattern can only be discerned after taking its statistical summation. However, coherent interference pattern for classical field is stationary. This striking feature can distinguish two kinds of interference mechanism.

The experimental results can be verified by the theoretical simulation. For simplicity, we consider a one-dimensional case. When a coherent plane wave $E_0(x_0)=E_0$ driving the interferometer, the object field $E_o$ and reference field $E_r$ at the output port can be calculated by Eq. (1.2.1)

$$E_o(x) = \int T(x_0) H(x, x_0; z_o, z_o) E_0 dx_0$$

$$= E_0 \exp(ikz_o)\sqrt{\frac{k}{i2\pi z_o}}\int T(x_0)\exp\left[\frac{ik(x-x_0)^2}{2z_o}\right]dx_0, \quad (4.1.1a)$$

$$E_r(x) = \int H(x,x_0;z_r,z_r)E_0 dx_0 = E_0\exp(ikz_r). \quad (4.1.1b)$$

When the temporal coherence $|z_r-z_o|/c<<\tau$ is satisfied, the coherent interference pattern is obtained to be

$$\langle E_r^*(x)E_o(x)\rangle = E_r^*(x)E_o(x) = I_0\sqrt{\frac{k}{i2\pi z_o}}\exp[ik(z_o-z_r)]\int T(x_0)\exp\left[\frac{ik}{2z_o}(x-x_0)^2\right]dx_0, \quad (4.1.2)$$

where $I_0 = E_0^*E_0$ is the intensity of the coherent beam. The coherent interference pattern is stationary.

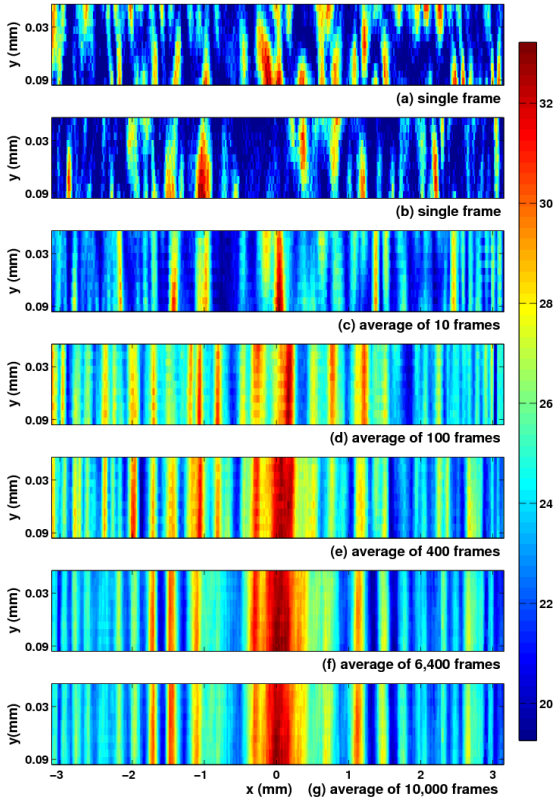

Fig. 4-2 Experimentally observed 2D interference patterns recorded by CCD$_1$. (a) and (b) are individual single frames; (c), (d), (e), (f) and (g) are averaged over 10, 40, 400, 6400 and 10000 frames, respectively. The color-bar shows the relative intensity (Zhang et al., 2010a).

Assuming that the coherent plane field is irregularly modulated by a rotating ground glass disk, $E_0(x_0)$ becomes a chaotic field satisfying spatial incoherence of Eq. (1.3.1). Again, when the temporal coherence is satisfied, the incoherent interference pattern at the output can be calculated to be

$$\langle E_r^*(x)E_o(x)\rangle = \int H^*(x,x_0';z_r,z_r)T(x_0)H(x,x_0;z_o,z_o)\langle E_0^*(x_0')E_0(x_0)\rangle dx_0'dx_0$$

$$= I_0\int T(x_0)H^*(x,x_0;z_r,z_r)H(x,x_0;z_o,z_o)dx_0. \quad (4.1.3)$$

Using Eq. (1.3.3), we obtain

$$\langle E_r^*(x)E_o(x)\rangle = I_0\exp[ik(z_o-z_r)]\int T(x_0)\exp\left[-\frac{ik}{2z_r}(x-x_0)^2\right]\exp\left[\frac{ik}{2z_o}(x-x_0)^2\right]dx_0$$

$$= I_0\exp[ik(z_o-z_r)]\int T(x_0)\exp\left[\frac{ik}{2Z}(x-x_0)^2\right]dx_0, \quad (4.1.4)$$

where the effective diffraction length $Z$ is defined by

$$\frac{1}{Z} = \frac{1}{z_o} - \frac{1}{z_r}. \tag{4.1.5}$$

We see that incoherent interference pattern represents the Fresnel diffraction integral of an object under the paraxial condition, the same as for coherent diffraction but with an effective diffraction distance $Z$ replacing the real one, $z_o$. Note that $Z$ is negative for $z_o > z_r$, designating the wavefront-reversal diffraction (it will be discussed later). As for $z_o = z_r$, the interference term does not vanish but the interference pattern is homogeneous.

For the coherent illumination, the intensity distribution of the reference wave is homogeneous whereas the intensity of the object wave itself exhibits the diffraction pattern

$$\langle E_o^*(x,t) E_o(x,t) \rangle = \frac{I_0 k}{2\pi z_o} \left| \int T(x_0) \exp\left[ \frac{ik}{2z_o}(x - x_0)^2 \right] dx_0 \right|^2. \tag{4.1.6}$$

In holography this pattern should be avoided by diminishing the intensity of the object wave. However, for the spatially incoherent illumination, the intensity distributions for both the object wave and reference wave are homogeneous. This is a significant advantage of incoherent interference in holographic applications.

To demonstrate the above theoretical explanation, Fig. 4-3 shows one-dimensional (1D) intensity patterns for incoherent interference (left) and coherent interference (right), respectively, where (a) and (b) are the interference fringes observed by $CCD_1$ and $CCD_2$, respectively. For a 50/50 beamsplitter $BS_2$, if one output field is $[E_r(x) + E_o(x)]/\sqrt{2}$ the other should be $[E_r(x) - E_o(x)]/\sqrt{2}$. Taking the difference and sum of the two output intensities gives, respectively, the net interference pattern and the intensity background, as shown in (c) and (d). The left and right interference fringes in Fig. 4-3(c) are observed in the same interferometer illuminated by the incoherent beam and coherent beam, respectively. Obviously, they have different fringe spacing, because in this experimental scheme the former has the effective diffraction length $Z=39.3$cm while latter has the diffraction length $z_o=16$cm. On the other hand, incoherent interference has the homogeneous intensity background of two arms in the left part of Fig. 4-3(d), while the coherent one has included the interference fringe of the object arm in the right part of Fig. 4-3(d), which is described by Eq. (4.1.6).

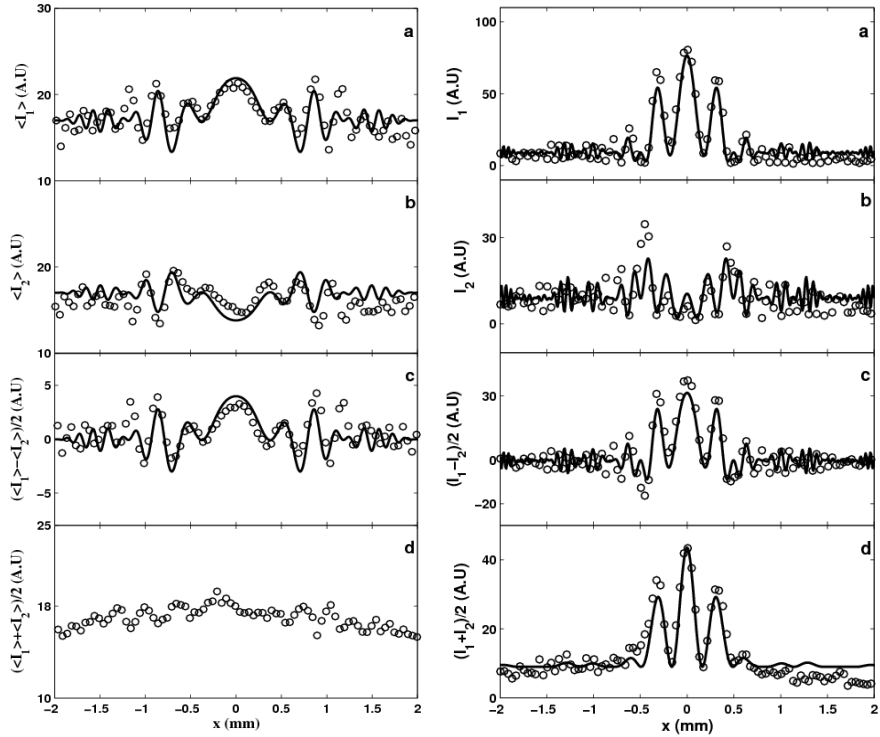

Fig. 4-3 Experimentally observed 1D interference patterns. Left and right parts are the patterns obtained using spatially incoherent and coherent light, respectively. (a) and (b) are the patterns registered by $CCD_1$ and $CCD_2$, respectively; (c) and (d) are their difference and summation, respectively. The patterns for the incoherent case are averaged over 10000 frames. Experimental data and theoretical simulation are given by open circles and solid lines, respectively (Zhang et al., 2010a).

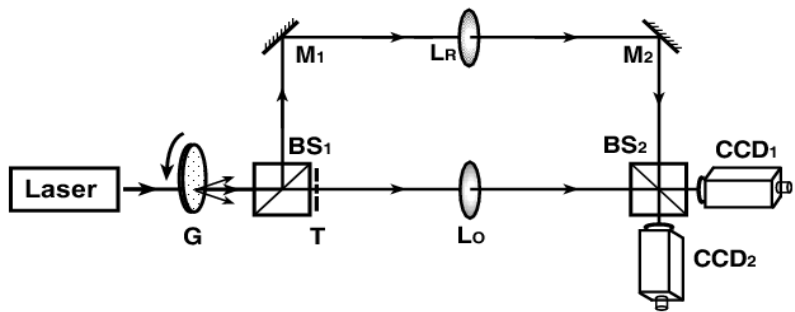

Fig. 4-4 Unequal-optical-path interferometer with lenses, which are inserted in the centre of the two arms. Both the input and output ports of the interferometer are located at the two focal planes of the lenses (Gao et al., 2009).

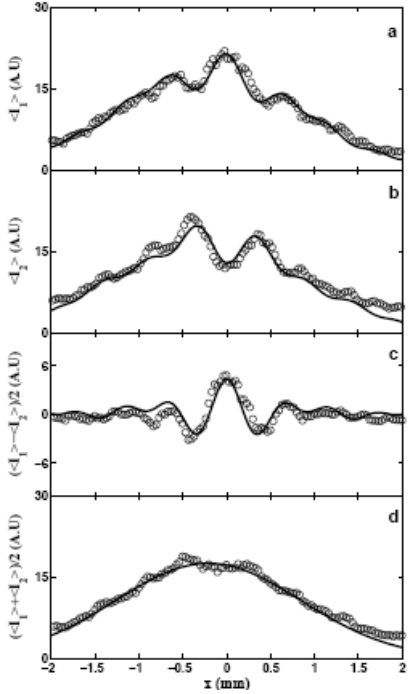

Fig. 4-5 Experimentally observed 1D interference patterns with spatially incoherent illumination in the scheme of Fig. 4-4. (a) and (b) are interference patterns (averaged over 10000 frames) registered by CCD$_1$ and CCD$_2$, respectively; (c) and (d) are their difference and summation, respectively. Experimental data and theoretical simulation are given by open circles and solid lines, respectively (Gao et al., 2009).

We can also use lens to perform Fourier transform for spatially incoherent illumination in the unequal-optical-path interferometer. As shown in Fig. 4-4, the previous interferometer is modified by inserting two lenses in the centre of the two arms (Gao et al., 2009a). Since a lens can perform Fourier transform of a field between its two focal planes, we position both the input and output ports of the interferometer at the two foci of the lenses. The impulse response function between the two focal planes for a lens of focal length $f$ is given by

$$F(x, x_0; f) = \sqrt{k/(i2\pi f)} \exp(i2kf)\exp(-ikxx_0/f). \qquad (4.1.7)$$

Let an object $T(x_0)$ close to the input port. In the incoherent case, similar to Eqs. (4.1.3) and (4.1.4), the interference term is obtained to be

$$\langle E_r^*(x)E_o(x)\rangle = I_0\int T(x_0)F^*(x,x_0;f_r)F(x,x_0;f_o)dx_0$$

$$= \frac{kI_0}{2\pi\sqrt{f_o f_r}}\exp[i2k(f_o - f_r)]\int T(x_0)\exp(-ikxx_0/f_{eff})dx_0, \qquad (4.1.8)$$

where $f_o$ and $f_r$ are the focal lengths of the lenses in the object arm and reference arm, respectively. The effective focal length $f_{eff}$ is defined as

$$\frac{1}{f_{eff}} = \frac{1}{f_o} - \frac{1}{f_r}. \qquad (4.1.9)$$

In the experiment the two lenses of focal lengths $f_o$=7.5cm and $f_r$=12cm, and a double-slit of slit width $b$=125μm and spacing $d$=310μm as the object are taken. The experimental results are plotted in Fig. 4-5. After eliminating the intensity background, Fig. 4-5(c) shows the Fourier spatial spectrum of the double-slit with the effective focal length $f_{eff}$=20cm.

If a coherent plane wave drives the same interferometer, the interference term is given by

$$E_r^*(x)E_o(x) = I_0\sqrt{f_r/f_o}\exp[i2k(f_o - f_r)]\delta(x)\int T(x_0)\exp(-ikxx_0/f_o)dx_0, \qquad (4.1.10)$$

where Dirac Delta function $\delta(x)$ comes from the focusing effect of the lens in the reference arm. This means that in the coherent illumination only the zero-frequency component of the Fourier spectrum of the object can be extracted in the interferometer. To obtain the Fourier spectrum of the object for the coherent interferometer, the lens in the reference arm has to be removed.

The above theoretical analysis and experimental demonstration tell us that, in an unequal-optical-path interferometer, spatially incoherent light is capable of performing holographic interference in a similar way to that of coherent light. Physically, the interference mechanisms of the two types of sources are different, so do the interference patterns. In the incoherent case, the interference pattern fluctuates irregularly in time, and a well-defined pattern can only be formed in the statistical summation. Moreover, the diffraction pattern of an object is dependent on the effective diffraction length, which is associated with the correlated-field diffraction of both the object and reference waves.

## 2. Incoherent interference in an equal-optical-path interferometer

In the interferometric schemes above, the two arms have different optical path lengths. So the light source is required to have much better temporal coherence, and it is possible only for a laser source. A quasi-monochromatic true thermal light source has a very short coherence time less than 0.1 nsec and fails in these schemes. From an application point of view, we must look for other schemes if we expect to employ a true thermal light source in the first-order incoherent interference. For this purpose, two arms of interferometer must have equal optical path lengths.

We have already pointed out in Sec. I-3 that incoherent interference does not exist in the original Gabor's holographic scheme, since both the object and reference waves travel in exactly the same diffraction configuration. In 2009, Zhang et al. proposed an equal-optical-path interferometer, which evades the complete cancellation of diffraction in two arms (Zhang et al., 2009a). As shown in Fig. 4-6, the two arms of the interferometer have the same length, $2f_o$. A lens $L_o$ of focal length $f_o$=19cm is set in the centre of the object arm, while in the other arm the beam travels freely to beamsplitter $BS_2$. Hence the diffraction through the lens in the object arm has a different configuration from that for free propagation in the other arm.

Let object $T(x)$ and the CCD cameras be placed at the two focal planes of lens $L_o$, which produces the Fourier transform of the object. Under the illumination of a coherent plane wave, the interference term given by Eq. (1.2.4) exhibits the Fourier spatial spectrum of the object

$$E_r^*(x)E_o(x) \propto \int T(x_0)\exp(-ikxx_0/f)dx_0 \equiv \tilde{T}(kx/f), \tag{4.2.1}$$

where $\tilde{T}(kx/f)$ is the Fourier transform of $T(x_0)$. However, when S is a spatially incoherent source satisfying Eq. (1.3.1), according to Eq. (1.3.2) we obtain the interference term to be

$$\langle E_r^*(x)E_o(x)\rangle \propto \int T(x_0)F(x,x_0;f)H^*(x,x_0;2f,2f)dx_0$$

$$\propto \int T(x_0)\exp\left[-\frac{ik}{4f}(x+x_0)^2\right]dx_0 \approx \exp\left[-\frac{ik}{4f}x^2\right]\tilde{T}[kx/(2f)], \tag{4.2.2}$$

which shows the Fresnel diffraction pattern of the object with a diffraction length $2f$, and the last step is valid in the far-field limit.

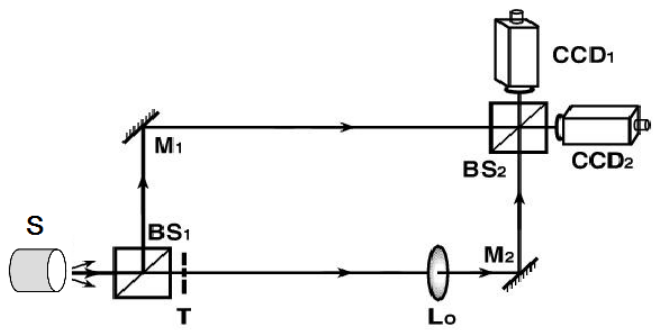

Fig. 4-6 An equal-optical-path interferometer: S is a thermal light source; a lens $L_o$ of focal length $f$ is set in the centre of the object arm, and both arms have the same length $2f$. (Zhang et al., 2009a)

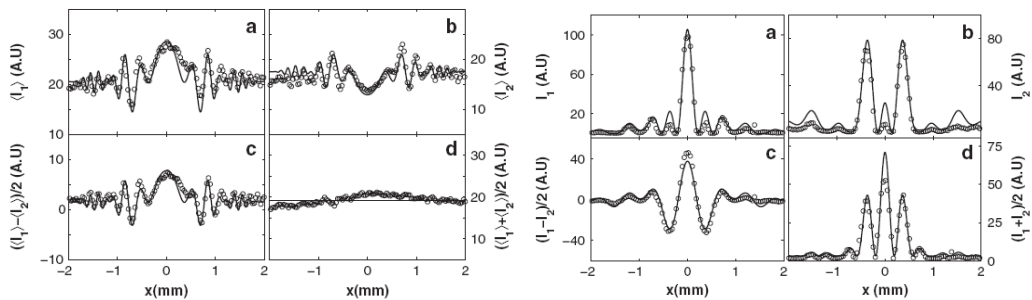

Fig. 4-7 Experimentally observed 1D interference patterns in the scheme of Fig. 4-6. Left and right parts are the patterns using spatially incoherent and coherent light, respectively. (a) and (b) are interference patterns (averaged over 10000 frames for the incoherent case) registered by $CCD_1$ and $CCD_2$, while (c) and (d) are their difference and summation, respectively. Experimental data are given by open circles and theoretical simulation by solid lines (Zhang et al., 2009a).

With the experimental scheme in Fig. 4-6, four types of sources are employed. Figure 4-7 shows 1D patterns where the left part corresponds to a He-Ne laser beam passing through a ground glass plate (the same as in Fig. 4-1) while the right part, the direct laser beam by removing the ground glass plate. We can see that the interference patterns in the left part are very similar to those in Fig. 4-3. As a matter of fact, these two incoherent schemes exhibit similar Fresnel diffraction with slightly different effective diffraction lengths, $Z=39.3$cm and $2f=38$cm. When the ground glass plate is removed, the coherent interference patterns are registered by $CCD_1$ and $CCD_2$, as shown in the right part of Figs. 4-7(a) and 4-7(b), respectively. They are the spatial Fourier spectra of the double-slit, consisting of two parts, $\tilde{T}(kx/f)$+c.c. and $|\tilde{T}(kx/f)|^2$. The former and the latter can be extracted by the difference and sum of (a) and (b), as shown in the right part of Figs. 4-7(c) and 4-7(d), respectively.

The third source is a Na lamp which emits true thermal light of wavelength 589.3 nm. The source has an illumination area of $10\times10$ mm$^2$. In this case, the interference pattern can be directly observed on the CCD screen, as shown in Fig. 4-8(b). This is due to the fact that the response time of the CCD camera is much longer than the time scale of the thermal light fluctuations, so that averaging has already taken place in a single CCD exposure. For comparison, Fig. 4-8(a) shows the 2D interference pattern corresponding to Fig. 4-7(a, left) for the pseudo-thermal light. The two sets of fringes are similar, but have slightly different spacings, due to the different wavelengths of the two sources. Finally, a pinhole of diameter 0.36 mm is set after the Na lamp to dispel the spatial incoherence. With this point-like source, a different interference pattern appears on the CCD screen, as shown in Fig. 4-8(c). Consequently, the present scheme is favorable for realizing incoherent interference using a true thermal source.

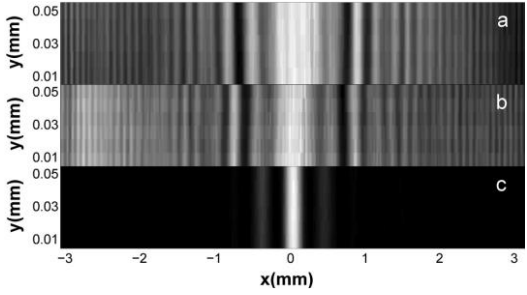

Fig. 4-8 Experimentally observed 2D interference patterns registered by $CCD_1$ in the scheme of Fig. 4-6: (a) with the pseudo-thermal light source, averaged over 10000 frames; (b) with a Na lamp of extended illumination area as the light source; (c) with a Na lamp followed by a pinhole as the light source (Zhang et al., 2009a).

The above experimental results tell us that spatially incoherent and coherent thermal sources with the same Na lamp produce different interference fringes in the same equal-optical-path interferometer. On the other hand, we can also expect that the incoherent interference pattern of Fig. 4-8(b) does not exist in a single exposure of coherence time of thermal light. However, the coherent interference pattern of Fig. 4-8(c) can exist in each moment: intensity fluctuation of the thermal light source does not destroy the interference pattern.

In coherent interference, two interfered coherent fields are not correlated, so the interference term records directly the diffraction of the object wave, as shown in Eq. (4.1.2). However, in the first-order incoherent interference schemes discussed above, we have just seen that the interference term can show well defined diffraction pattern of a double-slit in the ensemble average. As shown in Eq. (4.1.4), for example, the first-order interference term is related to the correlation of two incoherent diffraction fields, i.e. correlated-field diffraction (CFD). As a result the CFD pattern can be seen in intensity distribution but not intensity correlation. In both unequal- and equal-optical-path interferometers, CFD pattern can record complete information of an object. Incoherent holography technique proposed in the 1960s can also be understood with CFD, where incoherent light from an object experiences two different diffraction configurations, as shown in Eq. (2.4). Only when two interfered fields experience the same diffraction configuration, $h_r=h_o$, CFD is canceled exactly and the incoherent interference pattern is washed out as well.

## 3. Phase sensitive ghost imaging in first-order incoherent interference

First-order incoherent interference related to correlated-field diffraction (CFD) reveals more interesting phenomena than coherent interference. In Sec. III-6, we have seen that CFD in thermal light gives rise to ghost diffraction and ghost imaging, where the diffraction/imaging pattern is observed through intensity correlation measurement of two fields. The question we wish to address is whether similar imaging can be realized in first-order incoherent interference.

### 3.1 *Wavefront-reversal diffraction and lensless phase sensitive imaging in first-order incoherent interference*

As has already been pointed out in Sec. IV-1, effective diffraction length of correlated-field diffraction in an unequal-optical-path interferometer could be negative, designating wavefront-reversal diffraction. The wavefront-reversal diffraction can refer to optical phase conjugation, which was discovered in a nonlinear optical process, where a phase conjugate wave is created in the way of both time reversal and wavefront reversal with respect to incident wave.

Incoherent interference provides a way to reveal wavefront-reversal diffraction, despite the fact that the phase conjugation wave has not been created physically. Interestingly, lensless imaging as the intrinsic effect of phase conjugation in nonlinear optical process can also be found not only in second-order incoherent interference but also in first-order one.

In 2009, Zhang et al. (Zhang et al., 2009b) proposed other equal-optical-path interferometer which is a simple modification of the previous scheme in Fig. 4-1. As shown in Fig. 4-9, a transparent glass rod is inserted into the short arm, and the length of the rod is selected in such a way that the two arms of the interferometer have equal optical path length although their geometric lengths are different. Thus the interferometer is equal-optical-path one and does not require the driving beam to have better temporal coherence.

When a wave travels over a distance $z$ in a homogeneous material of refractive index $n$, the impulse response function is given by $H(x,x_0;nz,z/n)$ (see Eq. (1.3.3)), where the optical path length $nz$ is different from the diffraction length $z/n$. In successive free propagation through two regions of optical path length $n_j z_j$ and diffraction length $z_j/n_j$ ($j=1,2$), we obtain the impulse response function of the cascade diffraction

$$H(x,x_0;n_1z_1+n_2z_2,z_1/n_1+z_2/n_2)=\int H(x,x';n_2z_2,z_2/n_2)H(x',x_0;n_1z_1,z_1/n_1)dx'. \quad (4.3.1)$$

As an example, we calculate the impulse response function of the reference arm in the present interferometer. Let the geometric length of the reference arm is $z_r$ while the glass rod is of length $l$ and refractive index $n$, we obtain $H(x,x_0;Z,\bar{Z})$, where the optical path length is $Z=z_r-l+nl$ and the diffraction length is $\bar{Z}=z_r-l+l/n$.

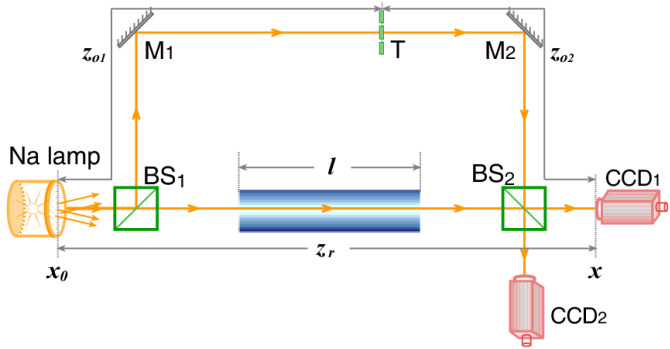

Fig. 4-9 An equal-optical-path interferometer, where a glass rod of length $l$ is placed in one arm, and T is an object in the other arm. The interferometer is illuminated by a sodium lamp. (Zhang et al., 2009b)

In the other arm, object $T(x)$ is placed at the position which is at a distance $z_{o1}$ from the source and $z_{o2}$ from either of the CCD cameras. The impulse response function is written as

$$h_o(x,x_0)=\int H(x,x';z_{o2},z_{o2})T(x')H(x',x_0;z_{o1},z_{o1})dx'. \quad (4.3.2)$$

For the perfect incoherence of the source, the interference term can be calculated to be

$$\langle E_r^*(x)E_o(x)\rangle = \int H^*(x,x_0;Z,\bar{Z})h_o(x,x_0)dx_0$$

$$\propto \exp[ik_0(z_o-Z)]\int T(x')\exp\left[\frac{ik_0}{2Z_{eff}}(x-x')^2\right]dx', \quad (4.3.3)$$

where $z_o=z_{o1}+z_{o2}$ is the length of the object arm, and the effective diffraction length $Z_{eff}$ is defined by

$$\frac{1}{Z_{eff}}=\frac{1}{z_{o2}}+\frac{1}{z_{o1}-\bar{Z}}. \quad (4.3.4)$$

As the equal-optical-path condition $z_o=Z$ is employed, the interferometer is capable of realizing incoherent interference with a true thermal light source. Under this condition the effective diffraction length becomes

$$Z_{eff} = \frac{z_{o2}(z_{o1} - \overline{Z})}{l(n - 1/n)}. \quad (4.3.5)$$

When the position of object is varied, the effective diffraction length can be positive or negative, corresponding to normal or wavefront-reversal diffraction, respectively. In particular, when $z_{o1} = \overline{Z}$, the effective diffraction length is $Z_{eff}$=0. This means that the diffraction will not occur, and the image of the object is reproduced with an equal size on the CCD screen.

In the experimental scheme of Fig. 4-9, the object and reference arms have lengths of $z_o$=41.8cm and $z_r$=33.8cm, respectively. When the length and refractive index of the glass rod are $l$=15.5cm and $n$=1.5163, respectively, the equal-optical-path condition $z_o$=Z of the interferometer is satisfied. Hence a true thermal light source, the Na lamp, can be used in the experiment. Moreover, the diffraction length of the reference arm can be calculated to be $\overline{Z} = 28.5\text{cm}$, so the imaging position in the object arm is $z_{o1}|_{img}$=28.5cm.

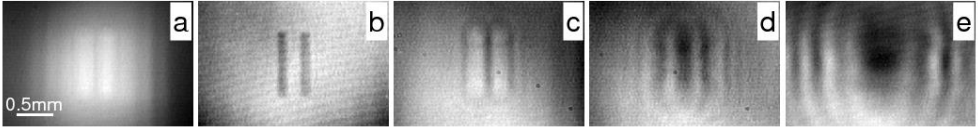

Fig. 4-10 Experimentally observed 2D interference patterns in the scheme of Fig. 4-9. [(a)-(e)]: the double slit is placed at the positions of $z_{o1}$=31.0, 28.5, 24.2, 20.0, and 10.6cm, respectively, where (b) is the image of the double slit (Zhang et al., 2009b).

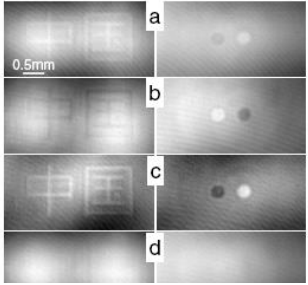

Fig. 4-11 Experimental results of lensless phase sensitive imaging of two objects. (a) and (b) are the images recorded by the two CCDs; (c) is the difference of (a) and (b) and (d) is their sum, respectively (Zhang et al., 2009b).

Figures 4-10(a)-(e) show the interference patterns for the double-slit placed, respectively, at the positions $z_{o1}$=31.0, 28.5, 24.2, 20.0, and 10.6cm, corresponding to the effective diffraction lengths $Z_{eff}$=2.0, 0, −5.7, −13.9, and −42.0 cm. The experiment has verified that the position for the object to be imaged (Fig. 4-10(b)) is at $z_{o1}$=28.5cm, which is equal to the diffraction length of the reference arm $\overline{Z}$. The regions left and right of this position correspond to positive and negative diffraction effects, respectively. Figures 4-11 shows the lensless phase sensitive imaging of two objects placed at the position of $z_{o1}$=28.5 cm from the source. Left column: an amplitude modulated object of two Chinese characters "China". Right column: a phase-modulated object of two holes with a phase difference of π. (a) and (b) are the images recorded by the two CCDs; (c) and (d) are their difference and sum, respectively. For the phase object of the two holes, the images exhibit distinct phase contrast: when one is bright, the other is dark. The imaging scheme is evidently phase sensitive.

3.2 *Phase sensitive ghost imaging with lens in first-order incoherent interference*

In 2009, Gao et al. (Gao et al., 2009b) reported an experiment of ghost imaging performed in first-order interference. As shown in Fig. 4-12, the interferometer in the experiment is equal-optical-path, very similar to that in Fig. 4-6, but a lens $L_2$ for imaging is set in the reference arm (path 2). The double-slit as object T and CCD camera are set at the two focal planes of lens $L_1$, so the equal-optical-path condition gives $z_1+z_2=z_0+2f_1$.

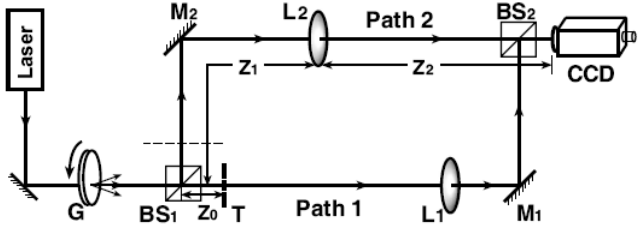

Fig. 4-12 Experimental setup of ghost imaging in first-order interference (Gao et al., 2009b).

Let $x_0$ and $x$ be the transverse positions in the source and detection planes, respectively. The impulse response functions of paths 1 and 2 are given by

$$h_1(x,x_0) = \frac{k_0}{i2\pi\sqrt{z_0 f_1}} \exp[ik_0(z_0+2f_1)] \int T(x') \exp\left[-\frac{ik_0 xx'}{f_1} + \frac{ik_0(x'-x_0)^2}{2z_0}\right] dx', \quad (4.3.6a)$$

$$h_2(x,x_0) = \sqrt{\frac{k_0}{i2\pi R}} \exp[ik_0(z_1+z_2)] \exp\left\{\frac{ik_0}{2R}\left[x_0^2\left(1-\frac{z_2}{f_2}\right) + x^2\left(1-\frac{z_1}{f_2}\right) - 2xx_0\right]\right\}, \quad (4.3.6b)$$

where $f_1$ and $f_2$ are the focal lengths of lenses $L_1$ and $L_2$, respectively. $z_0$, $z_1$, and $z_2$ have been marked in Fig. 4-12. $T(x)$ is the transmission function of the object, and $R=z_1 z_2(1/z_1+1/z_2-1/f_2)$. If the source is a coherent field satisfying Eqs. (1.2.3) and (1.2.4), one obtains two stationary patterns related to object superposed in the detection plane

$$\langle E_1^*(x)E_1(x)\rangle \propto \left|\int h_1(x,x_0)dx_0\right|^2 \propto |\tilde{T}(k_0 x/f_1)|^2, \quad (4.3.7a)$$

$$\langle E_1^*(x)E_2(x)\rangle + \text{c.c.} \propto \int h_1^*(x,x_0)dx_0 \times \int h_2(x,x_0)dx_0 + \text{c.c.}$$

$$\propto \tilde{T}^*(k_0 x/f_1)\exp\left[\frac{ik_0 x^2}{2(z_2-f_2)}\right] + \text{c.c.}, \quad (4.3.7b)$$

where $\tilde{T}$ is the Fourier transform of function $T$. Therefore coherent interference shows the Fourier spatial spectrum of the object. Consider now the perfect incoherent field satisfying Eq. (1.3.1). Under the ghost imaging equation

$$\frac{1}{z_1-z_0} + \frac{1}{z_2} = \frac{1}{f_2}, \quad (4.3.8)$$

one arrives the phase-sensitive image of the object in the incoherent first-order interference

$$\langle E_1^*(x)E_2(x)\rangle \propto \int h_1^*(x,x_0)h_2(x,x_0)dx_0$$

$$\propto T^*\left(-x\frac{z_1-z_0}{z_2}\right)\exp\left[\frac{ik_0 x^2}{2}\left(\frac{1}{f_2}-\frac{2}{f_1}\right)\frac{z_1-z_0}{z_2}\right]. \quad (4.3.9)$$

Moreover, the intensity distributions of two arms, $\langle E_1^*(x)E_1(x)\rangle$ and $\langle E_2^*(x)E_2(x)\rangle$, are homogeneous. In particular, let $f_1=2f_2$, with Eq. (4.3.8) and the equal-optical-path condition we obtain $z_1-z_0=z_2=f_1$. Equation (4.3.9) can be simplified as

$$\langle E_1^*(x)E_2(x)\rangle \propto T^*(-x). \quad (4.3.10)$$

This result shows phase sensitive imaging with equal size.

In the experimental scheme of Fig. 4-12, the incoherent source is formed by passing a He-Ne laser beam of wavelength 632.8 nm through a slowly rotating ground glass disk G. The focal lengths of the two lenses are $f_1$=15 cm and $f_2$=7.5 cm. The distances are as follows: $z_0$=1.5 cm, $z_1$=16.5 cm, and $z_2$=15 cm. Experimental results are shown in Fig. 4-13. Figure 4-13(a) shows the image of the double slit by an ordinary lens imaging system using incoherent light for comparison. By removing the ground glass disk G, an interference fringe of the double-slit appears in the screen, as shown in Fig. 4-13(b). Then the ground glass disk is inserted to form the incoherent interferometer, and the positive and negative images of the double slit are recorded in Figs. 4-13(c) and 4-13(d), respectively, depending on the path difference of the two arms by half a wavelength.

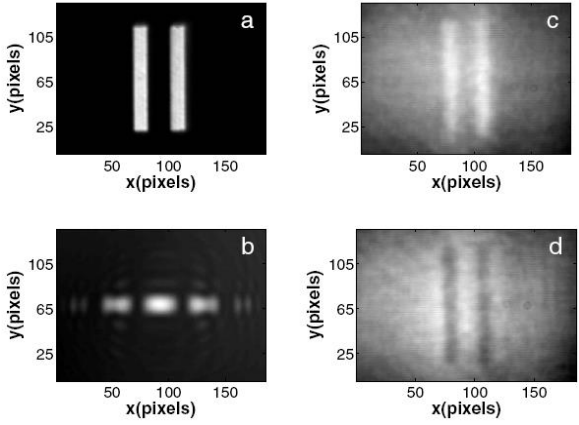

Fig. 4-13 Experimental results of the scheme of Fig. 4-12: (a) the image of the double-slit by ordinary lens imaging; (b) the interference fringe of the double-slit with coherent light; (c) and (d) the two images of the double-slit with incoherent light, averaged over 1000 frames (Gao et al., 2009b).

For comparison, we may set another bucket detector close to object T, which becomes the ghost imaging scheme in intensity correlation measurement (similar to Fig. 3-9b but object and imaging lens are not in one arm). The impulse response function of the object arm is now

$$h'_1(x, x_0) = \sqrt{\frac{k_0}{i2\pi z_0}} \exp(ik_0 z_0) \exp\left[\frac{ik_0(x-x_0)^2}{2z_0}\right] T(x) . \qquad (4.3.11)$$

The first-order field correlation function at two positions is given by

$$\langle E_1^*(x_1) E_2(x) \rangle \propto \int h'^*_1(x_1, x_0) h_2(x, x_0) dx_0 , \qquad (4.3.12)$$

where $E_1(x_1)$ and $E_2(x)$ are the fields at bucket detector and CCD, respectively. Under the imaging equation (4.3.8), one obtains

$$\left|\langle E_1^*(x_1) E_2(x) \rangle\right| \propto |T(x_1)| \delta\left(x_1 + \frac{z_1 - z_0}{z_2} x\right). \qquad (4.3.13)$$

According to Eq. (3.1.3) the intensity correlation measurement can record the modulus square of the first-order field correlation function. After the bucket detection, the image of the object is formed as

$$\int \left|\langle E_1^*(x_1) E_2(x) \rangle\right|^2 dx_1 \propto \left|T\left(-\frac{z_1 - z_0}{z_2} x\right)\right|^2 . \qquad (4.3.14)$$

The similarity between the two schemes is ascribed to the same correlated-field diffraction effect. However, phase sensitive imaging can be formed only in first-order incoherent interference. Therefore we may conclude that the correlation of two separated fields can exist for a single photon.

**V Nonlocal double-slit interference**

When a coherent optical field is divided into two parts by a beamsplitter, the two outgoing fields propagate without correlation. This is a well-known nature for coherence. As for a two photon entangled state, however, there is nonlocal correlation between two photons no matter how far they are apart. Experimental observation of such quantum correlation of two photons is an interesting and important work. A nonlocal double-slit consists of two different and spatially separated apertures, whose superposition defines a double slit. To reveal the nonlocal nature of two entangled photons, the nonlocal double-slit interference experiment was proposed and carried out with a two-photon entangled source in 1999 (Fonseca et al., 1999b). A few years later, it has been found that the similar nonlocal double-slit experiment can be performed with intensity correlation measurement of thermal light (Cao et al., 2007; Gao et al., 2008). These interference effects are of second-order ones due to two-photon measurement. In 2009, Gan et al. proposed a nonlocal double-slit experiment using first-order incoherent interference (Gan et al., 2009).

1. **Second-order nonlocal double-slit interference with a quantum entangled light source**

A two-photon entangled state can be generated by the spontaneous parametric down conversion (SPDC) process in a nonlinear crystal. As shown in Fig. 5-1, the signal and idler photon pairs generated in SPDC are scattered by two spatially separated apertures: none of them is a double-slit but their superposition at the same place forms a double-slit. The experimental results in Fig. 5-2 showed that the interference fringes appeared in two-photon coincidence measurements of scanning one detector while fixing the other. However, the individual signal and idler intensity profiles did not exhibit any fringe. The effect was attributed to the nonlocal correlation character of quantum entanglement.

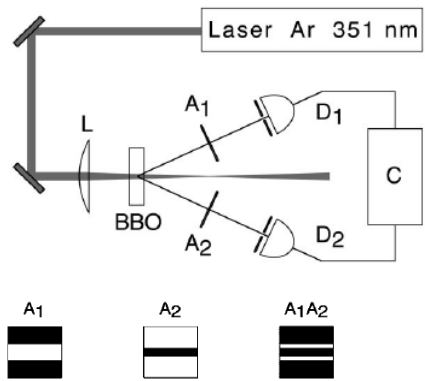

Fig. 5-1 Sketch of the nonlocal double-slit experiment with a two-photon entangled source. Bottom: detail of the apertures (Fonseca et al., 1999b).

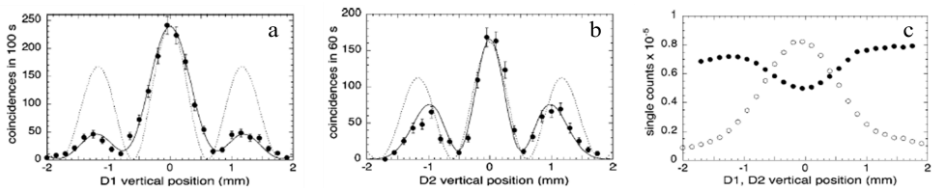

Fig. 5-2 Experimental results of the nonlocal double-slit interference with a two-photon entangled source. (a) and (b) Interference fringes are observed by the coincidence measurements of scanning one detector while fixing the other. (c) Single counts of each detector position (Fonseca et al., 1999b).

When the pump beam in a SPDC process has a transverse spatial spectrum of $C(p)$, the generated two-photon entangled state can be expressed as

$$|\psi_t\rangle = \int dp_1 dp_2 C(p_1 + p_2) a_1^\dagger(p_1) a_2^\dagger(p_2)|0\rangle = \int dx_1 dx_2 C(x_1)\delta(x_1 - x_2) a_1^\dagger(x_1) a_2^\dagger(x_2)|0\rangle, \quad (5.1.1)$$

where $p_i$ and $x_i$ are transverse momentum and position for photon $i$, and $C(x)$ is the Fourier transform of $C(p)$. For the ideal case, the pump beam is approximately assumed as a plane wave, it has $C(p)=\delta(p)$ and $C(x)=1$. This is the perfect two-photon entanglement that we have already discussed in Eq. (3.3.8). In this case, as we have pointed out in Sec. III-6, "The quantum entangled source behaves as a mirror,...", the two apertures will not form a nonlocal double-slit.

In the above experiment, the pump beam passes through a lens before driving the crystal. The plane wave approximation is no longer valid. Using Eq. (5.1.1), the two-photon amplitude of the entangled source is written as

$$\langle 0|E_{10}^{(+)}(x_1) E_{20}^{(+)}(x_2)|\psi_t\rangle \propto C(x_1)\delta(x_1 - x_2). \quad (5.1.2)$$

When two down-converted photons travel two different paths of the impulse response functions $h_1(x_1,x_0)$ and $h_2(x_2,x_0)$, the two-photon amplitude at the two detectors is obtained to be

$$\langle 0|E_1^{(+)}(x_1) E_2^{(+)}(x_2)|\psi_t\rangle = \int h_1(x_1, x_0) h_2(x_2, x_0') C(x_0) \delta(x_0 - x_0') dx_0$$

$$= \int h_1(x_1, x_0) h_2(x_2, x_0) C(x_0) dx_0. \quad (5.1.3)$$

In the experiment the two entangled photons are degenerate, $k_1=k_2\equiv k=k_p/2$, where $k_p$ is the wave number of the pump beam. By using Eq. (3.6.1), the impulse response function of each down-converted beam is calculated to be

$$h_j(x, x_0) = \frac{k}{i2\pi\sqrt{z_0 z}} \exp[ik(z_0 + z)] \exp\left[\frac{ik}{2}\left(\frac{x^2}{z} + \frac{x_0^2}{z_0}\right)\right]$$

$$\times \int A_j(x') \exp\left\{ik\left[\frac{z_0 + z}{2z_0 z} x'^2 - \left(\frac{x}{z} + \frac{x_0}{z_0}\right)x'\right]\right\} dx', \quad (j = 1,2) \quad (5.1.4)$$

where $A_j(x)$ is the transmission function of aperture $A_j$, and $z_0$ and $z$ are distances from aperture to crystal and detector, respectively. The pump beam is assumed as a Gaussian beam

$$C(x) = \exp[-ikx^2/(2f)], \quad (5.1.5)$$

where $f$ is the focal length of an effective lens. When the condition $z_0=2f$ is satisfied, the two-photon coincidence counting rate can be reduced to

$$\left|\langle 0|E_1^{(+)}(x_1) E_2^{(+)}(x_2)|\psi_t\rangle\right|^2 = \left|\int A_1(x') A_2(-x') \exp\left[ik_p\left(\frac{z_0 + z}{2z_0 z} x'^2 - \frac{x_1 - x_2}{2} \cdot \frac{x'}{z}\right)\right] dx'\right|^2. \quad (5.1.6)$$

In this way the two apertures can form a double-slit, $D(x)=A_1(x)A_2(-x)$. As a matter of fact, when a point source is distant $z_0$ from the double-slit, the coherent interference pattern is written as

$$I(x) = \left|\int D(x') \exp\left[ik\left(\frac{z_0 + z}{2z_0 z} x'^2 - x \cdot \frac{x'}{z}\right)\right] dx'\right|^2, \quad (5.1.7)$$

where $z$ is the distance between double-slit and observation plane. In comparison with the above coherent interference, the nonlocal double-slit interference of Eq. (5.1.6) can reveal two results. When the two detectors move oppositely, $x_1=-x_2=x$, the subwavelength interference happens. However, when either one detector is scanned while the other is fixed, the fringe is the same as the coherent case.

Equation (5.1.3) also shows that when a Gaussian beam (5.1.5) pumps the crystal, the quantum entangled source acts as a mirror plus a lens. For example in the Fonseca's experiment the two apertures are apart 2*f* from the crystal.

2. **Second-order nonlocal double-slit interference with a thermal light source**

The nonlocal double-slit experiment can also be performed with the intensity correlation measurement of thermal light. The experimental setup is shown in Fig. 5-3, where the two apertures are symmetrically placed from beamsplitter BS. The experiment result showed the interference fringes in Fig. 5-4 (Gao et al., 2008).

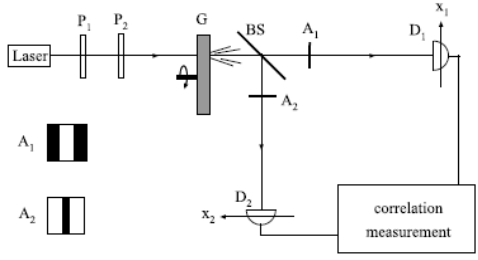

Fig. 5-3 Sketch of the experiment setup of the second-order nonlocal double-slit interference. P$_1$ and P$_2$ are a polarizer and a Glan prism, respectively. G and BS are a ground glass and a beamsplitter, respectively. A$_1$ and A$_2$ are two apertures placed at the symmetrical position with respect to the BS. D$_1$ and D$_2$ are CCD detectors (Gao et al., 2008).

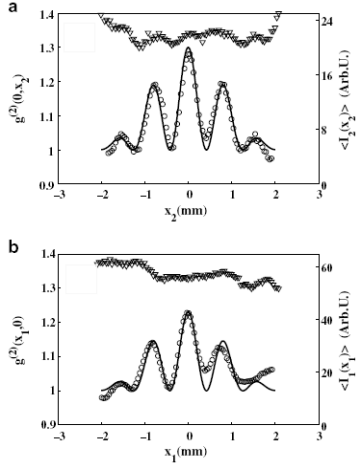

Fig. 5-4 Experimental data of intensity distributions (triangles) and normalized intensity correlations (circles). (a) Detector D$_2$ is scanned while detector D$_1$ is fixed; (b) vice versa. Numerical simulation of a typical double-slit interference fringe is shown by solid lines (Gao et al., 2008).

The above experiment verifies that the quantum entanglement is not necessary in the nonlocal interference effect. As a matter of fact, a nonlocal double-slit can be easier formed for the thermal light correlation—it does not need focused incident beam. The intensity correlation of thermal light measures the first-order field correlation function

$$\left|\left\langle E_1^*(x_1)E_2(x_2)\right\rangle\right|^2 = \left|\int h_1^*(x_1,x_0)h_2(x_2,x_0)\mathrm{d}x_0\right|^2$$
$$= \left|\int A_1^*(x')A_2(x')\exp\left[(\mathrm{i}k/z_2)(x_1-x_2)x'\right]\mathrm{d}x'\right|^2, \qquad (5.2.1)$$

where Eq. (5.1.4) has been applied. When $A_1^*(x')A_2(x')$ forms a nonlocal double-slit, the above equation shows the lensless Fourier transform of the double-slit function, similar to Eq. (5.1.6) in the far field limit for the entangled light case.

3. **First-order nonlocal double-slit interference with a thermal light source**

Similar to ghost imaging through single photon interference in subsection 3.2 of Sec. IV, the nonlocal double-slit effect can be performed through first-order incoherent interference. The experimental setup shown in Fig. 5-5 is similar to Fig. 4-6 but with two apertures replacing the real double-slit. The distances from each aperture to BS$_1$ and charge-coupled device (CCD) camera are $z_0$=1 cm and $z_1$=38 cm, respectively. The lens L of focal length $f$ =19 cm is set in the middle of A$_2$ and the CCDs. Figure 5-6 shows the interference patterns in the intensity distribution observed by individual CCDs. Similar to the left part of Fig. 4-7, they are the double-slit interference patterns in the interferometer.

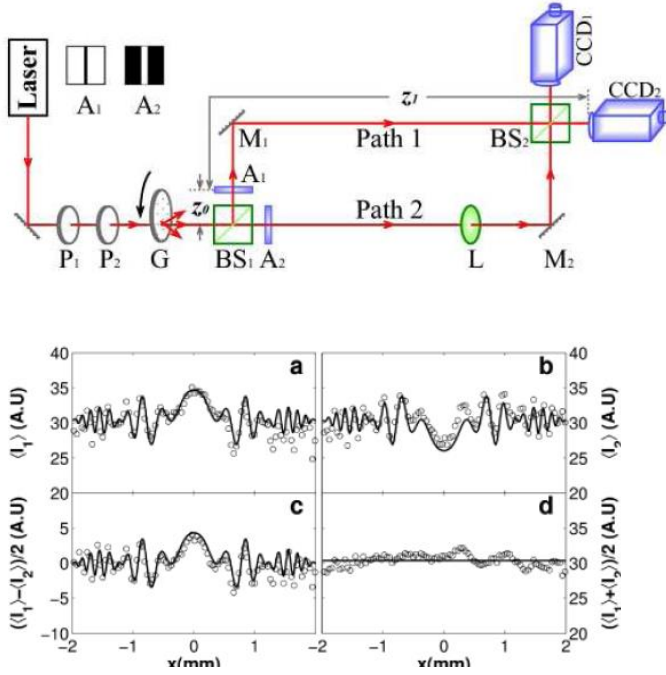

Fig. 5-6 Experimental data of intensity distributions (circles). (a) and (b) are the interference patterns registered by CCD$_1$ and CCD$_2$, respectively; (c) and (d) are their difference and summation, respectively (Gan et al., 2009).

In the interferometer the impulse response functions for the two paths are given by

$$h_1(x,x_0) = \frac{k}{\mathrm{i}2\pi\sqrt{z_0 z_1}} \exp[\mathrm{i}k(z_0+z_1)] \int A_1(x_1) \exp\left[\frac{\mathrm{i}k}{2z_0}(x_1-x_0)^2 + \frac{\mathrm{i}k}{2z_1}(x-x_1)^2\right] \mathrm{d}x_1, \quad (5.3,1a)$$

$$h_2(x,x_0) = \frac{k}{\mathrm{i}2\pi\sqrt{z_0 f}} \exp[\mathrm{i}k(z_0+2f)] \int A_2(x_2) \exp\left[\frac{\mathrm{i}k}{2z_0}(x_2-x_0)^2 - \frac{\mathrm{i}k}{f}xx_2\right] \mathrm{d}x_2. \quad (5.3,1b)$$

For the spatial incoherent source, the interference term is calculated to be

$$\langle E_1^*(x)E_2(x) \rangle \propto \int h_1^*(x,x_0)h_2(x,x_0)\mathrm{d}x_0 \propto \int A_1^*(x_0)A_2(x_0) \exp\left[-\frac{\mathrm{i}k}{4f}(x+x_0)^2\right] \mathrm{d}x_0, \quad (5.3.2)$$

where $z_1$=2$f$ has been applied. When the two apertures define a double-slit, this equation is exactly the same as Eq. (4.2.2).

The nonlocal double-slit interference experiments reveal intuitively quantum correlation and classical correlation effects, respectively existing in a two-photon entangled source and a thermal light source. We see again that for the thermal light source the nonlocal interference can be attributed to a single photon correlation. The similar scheme of nonlocal interference has been expanded to the nonlocal imaging with thermal light, which can be utilized in invisibility cloak and illusion phenomena (Gan et al., 2009; Zhang et al., 2010b).

## VI Summary

We have reviewed a number of interference experiments with spatially incoherent light sources. Incoherent interference can be generally defined as the interference for a spatially incoherent optical field where conventional coherent interference does not occur. A thermal light source and a two-photon entangled source can both be employed in incoherent interference experiments, which pertain to classical interference and quantum interference, respectively. For a classical thermal light source, intensity correlation manifests the first-order field correlation of two positions. As for a two-photon entangled source, the quantum correlation of two photons signifies biphoton coherence.

In first-order incoherent interference there is an explicit relation between coherent and incoherent effect. In a large-area thermal light source, each point-source creates its own instantaneous interference sub-pattern provided the temporal coherence is satisfied—this is a coherent effect although the sub-pattern as a whole fluctuates in time. An incoherent interference pattern can be formed by the summation of all the sub-patterns.

In the classical regime, coherent and incoherent interference phenomena have different characteristics. When the light to be interfered is sufficiently strong so that the shot noise can be neglected, a coherent interference pattern can appear in each single exposure, whereas an incoherent interference pattern can only be formed in the statistical summation.

Incoherent interference can be carried out through both single-photon measurement (for first-order effects) and two-photon correlation measurement (for second-order effects). To observe first-order incoherent interference, however, a well-designed interferometer in which two arms have different diffraction configurations is necessary.

Very interestingly, ghost imaging, lensless imaging and nonlocal double-slit experiments can also be observed in a first-order interference setup with thermal light. In these schemes, a single photon travels simultaneously two different paths and experiences nonlocal correlation if no measurement occurs along the way. The correlation information can be extracted in the interference plane through single-photon measurements. In this sense we may conclude that *a single photon can correlate with itself*.

**Acknowledgements**

I thank L. A. Wu and L. A. Lugiato for helpful discussions.